\documentclass[a4paper,fleqn]{cas-dc}

\usepackage[authoryear,longnamesfirst]{natbib}
\usepackage{titlecaps}
\Addlcwords{a}
\Addlcwords{an}
\Addlcwords{the}
\Addlcwords{and}
\Addlcwords{but}
\Addlcwords{or}
\Addlcwords{for}
\Addlcwords{nor}
\Addlcwords{in}
\Addlcwords{of}
\Addlcwords{on}
\Addlcwords{at}
\Addlcwords{to}
\Addlcwords{by}
\Addlcwords{with}
\Addlcwords{as}

\usepackage{fvextra}
\fvset{breaklines, breaksymbolright={}, breaksymbolleft={}}

\usepackage{adjustbox}
\usepackage{float}
\usepackage[section]{placeins}
\usepackage{caption}
\usepackage{longtable}

\begin{document}
\let\WriteBookmarks\relax
\def\floatpagepagefraction{1}
\def\textpagefraction{.001}

\shorttitle{LLMs' Misrepresentation of American Climate Opinions}

\title [mode = title]{How Large Language Models Misrepresent American Climate Opinions}

\author[1]{Sola Kim}
\cormark[1]
\cortext[1]{Corresponding author. ORCID: 0000-0002-2293-9348}
\affiliation[1]{{School of Sustainability, Arizona State University, 777 E University Dr, Tempe, AZ 85281, USA}}

\author[2]{Jieshu Wang}
\affiliation[2]{{Department of Technology and Society, Stony Brook University, 1432 Computer Science, Stony Brook, NY 11794, USA}}

\author[1]{Marco A. Janssen}
\author[1,3]{John M. Anderies}

\affiliation[3]{{School of Human Evolution and Social Change, Arizona State University, 900 S Cady Mall, Tempe, AZ 85281, USA}}

\begin{abstract}
 \relax Large language models are increasingly used to simulate public opinion, standing in for survey respondents as ``silicon samples.'' Whether they faithfully represent U.S. climate opinion, particularly at the intersections of race, gender, and ideology, remains largely untested. We prompted six large language models with the profiles of 978 respondents to a nationally representative U.S. climate opinion survey and compared the generated responses with those respondents' own answers across 20 questions. The models compress the diversity of American climate opinion, pulling predictions toward the population mean: less-concerned groups are predicted as more concerned, more-concerned groups as less, and the variance of predicted responses falls, on average, 29\% below that of human responses. The compression is intersectional. The models apply one gender assumption, predicting women as more pro-climate than men in every racial group. The assumption matches the direction of the human pattern for White, Hispanic, and Other Americans and runs the wrong way for Black Americans. We claim the direction of that difference rather than its size, since it rests on small cells at the ideological extreme. An audit of gender alone would report that the models understate the gender gap, and would miss that the error changes direction by race, which is what single-variable audits cannot detect. What is at stake is not accuracy alone but representation: whose views a synthetic sample renders legible. 
\end{abstract}


\begin{highlights}

\item  Six LLMs simulated the climate-change opinions of 978 U.S. survey respondents 
\item  The models' answers varied 29\% less than real people's, flattening disagreement 
\item  All six models assumed women care more about climate than men, in all racial groups 
\item  The assumption fits White, Hispanic and Other groups, but misfires for Black ones 
\item  Testing gender or race alone hides the misfire; only both together reveal it 

\end{highlights}


\shortauthors{Kim et al.}

\maketitle

\section{Introduction}
Federal agencies are deploying AI tools to analyze thousands of public comments on environmental regulations \citep{officeofmanagementandbudget2024FederalAI2024}. As of 2024, at least seven federal agencies use such tools, including Department of Energy's CALLM (Climate Action through Large Language Models) for climate resilience planning and Department of Interior's PCAT (Pragmatic Context Assessment Tool), which processes millions of public comments submitted under the National Environmental Policy Act. Meanwhile, researchers increasingly use large language models, AI systems such as ChatGPT trained to predict and generate human language. They use them to create what they call "silicon samples": simulated survey respondents that can rapidly and inexpensively approximate patterns in public opinion at scale \citep{argyleOutOneMany2023, schoeneggerWisdomSiliconCrowd2024}.  The two applications differ in a way that matters for what follows. Summarizing public comments processes text that people wrote; constructing a silicon sample generates responses that no one gave. This study examines the second application. Accuracy becomes consequential as these tools shape how public opinion is understood, and especially so for communities whose voices are already marginalized, where distortions in representation could compound existing inequities.

This question matters with particular urgency for climate change, where public opinion is contested and diverse \citep{druckmanEvidenceMotivatedReasoning2019, hornseyMetaanalysesDeterminantsOutcomes2016}. Climate attitudes appear to be shaped by a complex interplay of factors, including personal experience, affect, cultural values, and trust in science \citep{weberWhatShapesPerceptions2016, broschAffectEmotionsDrivers2021, leiserowitzClimateChangeRisk2006a}, many of which may not be captured by demographic profiles alone \citep{bergquistMetaanalysesFifteenDeterminants2022}. Political ideology in particular tends to cut across demographic lines, shaping how individuals interpret the same evidence \citep{druckmanEvidenceMotivatedReasoning2019, czarnekRightwingIdeologyReduces2021}; meta-analytic work suggests that values, ideologies, and political affiliation are among the strongest predictors of climate belief across nations \citep{hornseyMetaanalysesDeterminantsOutcomes2016}, and worldviews such as free-market ideology and social dominance orientation appear to drive motivated appraisals of climate science \citep{hornseyRoleWorldviewsShaping2021a}. Risk perceptions also vary not only between but within racial and ethnic groups \citep{maciasEnvironmentalRiskPerception2016}. This complexity is compounded by the fact that climate impacts fall disproportionately on marginalized communities, reflecting what scholars term "intersecting vulnerabilities": combinations of race, gender, and income that together shape how different groups experience climate risk \citep{kaijserClimateChangeLens2014, amorim-maiaIntersectionalClimateJustice2022, sultanaCriticalClimateJustice2022a, crenshawMappingMarginsIntersectionality1991}. Misrepresenting the views of these communities therefore risks compounding, rather than redressing, the inequities they already face \citep{eriksenAdaptationInterventionsTheir2021, chuInclusiveApproachesUrban2016}.

A growing body of research now examines whether large language models accurately represent different demographic groups \citep{gallegosBiasFairnessLarge2024, sinacolaLlmsVirtualUsers2025a}. Studies using LLMs as simulated survey respondents have found that these models can approximate aggregate response distributions and reproduce broad patterns in public opinion \citep{argyleOutOneMany2023, rafikovaChatGPTResearchProxy2026}.  Early work found that instruction-tuned models align poorly with several U.S. demographic groups and that steering toward a group only partly closes the gap \citep{santurkarWhoseOpinionsLanguage2023}. For climate specifically, ChatGPT has shown a pro-environmental orientation across experiments \citep{hartmannPoliticalIdeologyConversational2023}, a default that could push predictions toward higher climate concern than some groups actually hold.  These models also tend to produce responses that are more uniform than those of real populations, often capturing broad averages while understating how much people within the same group differ from one another \citep{bisbeeSyntheticReplacementsHuman2024, kaiserSimulatingHumanOpinions2025, aroraAIHumanHybrids2025}.

American climate opinion presents a useful test case for these concerns. In the United States, ideology and partisanship are among the strongest predictors of climate attitudes, often outweighing demographic variables such as age, gender, and education \citep{hornseyMetaanalysesDeterminantsOutcomes2016, dunlapWideningGapRepublican2008}. This partisan divide has deepened over time \citep{eganClimateChangeUS2017, mccrightPoliticizationClimateChange2011}, driven in part by elite cues and media environments \citep{brulleShiftingPublicOpinion2012}. The pattern is distinctly American; comparable ideological divides are more modest in Western Europe \citep{mccrightPoliticalIdeologyViews2016}. Opinions also diverge within racial groups by ideology and gender \citep{benegalRaceEthnicitySupport2022, stewartMappingRacialEthnic2024}. This structure suggests that LLMs conditioned on demographic profiles alone may miss the ideological axis that most shapes climate attitudes in the American context. Yet existing benchmarks, while now available for climate opinion, tend to simplify attitudes into a denier-believer binary, potentially overlooking this diversity \citep{kurfaliClimateEvalComprehensiveBenchmark2025}.

In the domain of climate opinion, few studies have examined whether such misrepresentation may vary across intersecting demographic categories. \citet{leeCanLargeLanguage2024} compared large language model predictions to a nationally representative U.S. climate opinion survey, finding that models underestimated Black Americans' climate beliefs relative to human responses. Yet this analysis examined demographic group accuracy without investigating whether prediction mismatches interact across intersecting identities: whether, for example, an AI model makes different types of mismatches when representing Black women versus Black men versus White women. Understanding these intersectional patterns matters because climate risk and vulnerability may vary along intersectional lines \citep{kaijserClimateChangeLens2014, amorim-maiaIntersectionalClimateJustice2022}. If LLM misrepresentations differ across intersecting identities, single-variable audits that examine race or gender in isolation may fail to detect them.

 To address this gap, our study asks: how do large language models represent demographic and intersectional patterns in U.S. climate change opinions? We prompted six large language models from different providers with demographic profiles describing 978 actual survey respondents and compared the AI-generated responses to the respondents' actual answers. Rather than measuring simple classification accuracy, we quantified the gap between model and human responses. 

Climate opinion has remained largely unexamined on this point. Our analysis confirms, now in the climate domain, the compression that prior work documented in other survey settings; the central contribution is to show that this compression is structured by intersecting identities: prediction mismatches do not shrink variance uniformly but reverse direction across combinations of race, gender, and ideology, so that a single demographic assumption overestimates some groups while underestimating others. Since offsetting mismatches cancel when race, gender, or ideology is examined alone, they escape the single-variable audits that dominate current practice. As these tools increasingly stand in for public opinion, such patterns could leave different intersectional groups misrepresented in the very settings meant to include them.

Our approach requires acknowledging what LLMs are and are not. These systems generate text by predicting likely continuations based on statistical patterns in training corpora. They do not model human cognition or access respondents' actual beliefs. When we prompt an LLM with a demographic profile, the output reflects associations encoded during training, which may overrepresent certain populations or perspectives while underrepresenting others.  We deliberately condition on demographics alone, excluding attitudinal covariates such as environmental group membership, prior climate policy engagement, or prior policy awareness. Two considerations motivate the choice. First, such covariates are close correlates of climate opinion itself, so supplying them would let the models read the outcome off a near-proxy and mask the demographic priors we set out to expose. Second, conditioning on demographics alone is the minimal specification standard in silicon-sample work \citep{argyleOutOneMany2023}, which isolates those priors when little else is given. Demographics, even with ideology, do not fully determine climate attitudes \citep{bergquistMetaanalysesFifteenDeterminants2022}, so some prediction gap is expected on design grounds; but an information limit would blur mismatches rather than reverse their sign across race and gender, so it would not produce the intersectional reversal we report.  These choices bound what we can claim: we identify systematic patterns in how LLMs represent intersectional groups relative to human survey data, not universal truths about AI bias. Whether these patterns stem from training data composition, model architecture, or emergent properties of scale remains beyond our scope. Yet the representational distortions themselves may be consequential regardless of origin.

\section{Intersectionality and Algorithmic Bias}

 Two bodies of scholarship inform this study: intersectionality and the study of algorithmic bias. Each arrived independently at the same conclusion, that harm at the intersection of social categories is not the sum of harms along each category alone. Intersectionality states it first. \citet{crenshawDemarginalizingIntersectionRace1989} named the problem in antidiscrimination law, where doctrine treated race and sex as separate and mutually exclusive, forcing Black women to press claims either as Black people or as women and leaving them unrecognizable as a distinct class. The erasure was structural. Legal and political frameworks organize around dominant reference points, so that feminist argument drew on the experience of white women and antiracist argument on the experience of Black men, and harms falling between them became difficult to name, recognize, or remedy \citep{crenshawDemarginalizingIntersectionRace1989, crenshawMappingMarginsIntersectionality1991}. Intersectionality is therefore an analysis of power, not an addition of categories \citep{collinsWhatIntersectionality2016}, and it resists reduction to a list of identities \citep{kaijserClimateChangeLens2014}. Climate scholarship bears this out on two fronts: intersectional analysis can reverse conclusions drawn from sex-disaggregated studies of climate vulnerability \citep{thompson-hallEnhancingExpandingIntersectional2016, djoudiDichotomiesGenderIntersecting2016a}, and American climate attitudes vary within racial groups by gender and ideology \citep{benegalRaceEthnicitySupport2022, stewartMappingRacialEthnic2024}. Harms organized around a dominant reference point are difficult to recognize, and an instrument built on that reference point cannot detect them. 

 Machine learning arrived at the same claim in its own vocabulary. \citet{kearnsPreventingFairnessGerrymandering2018} showed that a classifier fair on each protected attribute separately can be badly unfair on subgroups defined by their combination, a failure they term fairness gerrymandering. Marginal fairness does bound intersectional fairness, but too loosely for the guarantee to mean much \citep{molinaBoundingApproximatingIntersectional2022}. The canonical demonstration is \citet{buolamwiniGenderShadesIntersectional2018}: commercial gender classifiers were substantially less accurate for darker-skinned women than for lighter-skinned men, a disparity that aggregate accuracy conceals. Language models show the same structure. Intersectional biases emerge that their constituent categories do not predict \citep{guoDetectingEmergentIntersectional2021}, generative models prompted with occupations reproduce stereotypes most strongly at intersections \citep{kirkBiasOutoftheBoxEmpirical2021}, and comparable patterns appear in employment screening and across combinations of demographic attributes \citep{wilsonGenderRaceIntersectional2025, omarSociodemographicBiasesMedical2025, arzaghiUnderstandingIntrinsicSocioeconomic2025, goharSurveyIntersectionalFairness2023}. Auditing every subgroup is computationally hard, and single-variable audits remain the convention. Misclassification, however, is not the only harm at stake. A model that answers in a respondent's place does not classify them; it represents them. 

 Read together, the two accounts generate a specific expectation for language models. A model trained on text produced by a culture that organizes gender around a dominant reference point should learn a gender assumption fitted to that reference group, and should misrepresent whichever group departs from it. The doctrine \citet{crenshawDemarginalizingIntersectionRace1989} described could recognize \emph{Black} and it could recognize \emph{woman}, but not \emph{Black woman}; the question is whether a model prompted with a demographic profile does any better. Should it fail, the harm would not be a stereotype with a valence: an assumption that overstates a group's climate concern is no more accurate for being flattering. The harm would be erasure of a distinct position, an external definition substituted for the group's account of itself, which \citet{collinsPowerSelfDefinition1990} identifies as the displacement of self-definition. Existing evidence points that way. Language models prompted with demographic identities produce out-group portrayals rather than in-group self-descriptions \citep{wangLargeLanguageModels2025a}, and generative models systematically omit and stereotype minoritized groups \citep{shiehIntersectionalBiasesNarratives2026}. Algorithms do not merely record social conditions but help normalize them \citep{hoffmannWhereFairnessFails2019}. A model standing in for absent respondents produces knowledge about who those respondents are, rather than allocating a good among them. The harm at stake is therefore representational, and detecting it requires estimating how demographic effects interact rather than how each behaves alone.

\section{Methods}\label{sec:methods}

\subsection{Research Design}

This study employed an observational-computational comparison design. We used observational data from a nationally representative survey of U.S. adults to establish baseline human climate change opinions across demographic groups. We then conducted a computational experiment systematically manipulating demographic characteristics in prompts to six large language models, generating simulated responses that we compared to the human baseline. This design enabled us to examine whether and how LLMs represent demographic and intersectional patterns in climate opinions.

 The design is intersectional in a specific and limited sense. Following \citet{mccallComplexityIntersectionality2005}, we adopt an intercategorical approach: we take existing demographic categories provisionally, rather than as natural kinds, in order to document relations of inequality across the groups they mark, comparing, for example, how the gender gap in climate opinion differs between Black and White respondents. \citet{else-questIntersectionalityQuantitativePsychological2016} caution that no estimator is intersectional on its own, and that interaction terms acquire that status only when categories are framed in relation to power and results are interpreted accordingly. Three cautions follow. \citet{bowlegWhenBlackLesbian2008} supplies two: additive instruments tend to produce additive results, since presenting identity as a list of separable attributes invites separable processing, and null interactions are not evidence that structural inequality is absent. The third is that intersectional cells thin combinatorially, leaving estimates at the extremes unstable \citep{fouldsIntersectionalDefinitionFairness2020}. Sparsity is intrinsic to the approach, but no cell of a single respondent can carry a claim. 

\subsection{Human Survey Data and Sample Selection}

We used data from the Climate Change in the American Mind (CCAM) survey \citep{leiserowitzClimateChangeAmerican2020}, a nationally representative study conducted biannually since 2008. CCAM uses the Ipsos KnowledgePanel®, a probability-based online panel that uses address-based sampling to cover virtually all non-institutional U.S. households, including those without prior internet access. We analyzed the December 2024 wave, the most recent available at the time of data collection, to compare current human attitudes with contemporary LLM outputs.

The December 2024 wave included 1,013 respondents, from which we randomly selected 1,000 to match historical CCAM wave sizes and allow future testing of additional waves. We then excluded 22 cases with missing political ideology. The LLM prompt requires each respondent's self-reported ideology to construct the demographic persona (e.g., "You consider yourself somewhat conservative"), and ideology is the strongest predictor of American climate opinion \citep{hornseyMetaanalysesDeterminantsOutcomes2016, dunlapWideningGapRepublican2008}; respondents without this variable could neither be accurately prompted nor included in the intersectional analysis. The remaining 978 respondents formed the analytical sample. The excluded respondents were disproportionately female, Black, lower-income, and from the South (Table~\ref{tab:excluded_vs_retained}). While this exclusion is small (2.2\%), it may slightly underrepresent these groups in the analytical sample.

The sample included 508 males (51.9\%) and 470 females (48.1\%). Age ranged from 18 to 98 years ($M = 51.93$, $SD = 18.74$). The racial/ethnic composition consisted of 648 White non-Hispanic (66.3\%), 105 Black non-Hispanic (10.7\%), 135 Hispanic (13.8\%), and 90 Other non-Hispanic (9.2\%) respondents. Political ideology was distributed across very liberal (8.2\%), liberal (20.3\%), moderate (41.5\%), conservative (20.9\%), and very conservative (9.1\%) categories. Educational attainment ranged from less than high school (32.8\%) to bachelor's degree or higher (42.4\%), with some college representing 24.7\%. Household income categories included less than \$50,000 (20.45\%), \$50,000--\$99,999 (28.7\%), and \$100,000 or more (50.8\%). Regional distribution comprised the Northeast (17.7\%), West (23.6\%), Midwest (21.1\%), and South (37.6\%). Religiosity was categorized as religious (88.1\%) and not religious (11.9\%). Table~\ref{tab:demographic-distribution} presents the demographic composition of our final sample.

\subsection{Prior Research Using CCAM Data}

Prior research using CCAM data has examined how LLMs represent public climate opinions. \citet{leeCanLargeLanguage2024} investigated whether GPT-3.5 and GPT-4 could accurately predict individual beliefs using the 2017 and 2021 waves, finding that models systematically underestimated Black Americans' climate beliefs. However, their analysis examined each demographic group separately rather than testing whether mismatches interact across identities. Our study extends this work in three ways.

First, we examine prediction mismatches across intersecting demographic identities by analyzing how mismatches differ when race interacts with other demographic variables (gender, ideology, age, education, income, region, religiosity), rather than analyzing each subgroup separately. Consider the difference: comparing Black versus White average accuracy could obscure cases where women within a group are overestimated in the pro-climate direction while men are underestimated. Such patterns become invisible to analyses that do not separately examine gender within racial groups. Full interaction results appear in Table~\ref{tab:interaction_race}; we focus here on statistically significant findings.

Second, we measure misrepresentation as a continuous gap between LLM and human responses, rather than categorical accuracy, capturing both the direction and magnitude of misrepresentation.

Third, we restrict model conditioning to demographic variables alone, deliberately excluding issue-relevant covariates such as climate discussion frequency. While such covariates may improve prediction accuracy, including them can create circularity and does not reflect real-world deployment conditions where such data is typically unavailable.

\subsection{Large Language Models}

We tested six widely-used LLMs selected to represent diverse model architectures: GPT-5 (gpt-5-2025-08-07), GPT-4o (gpt-4o-2024-08-06), Llama-3 8B, Mistral 7B, Phi-4 14B, and Gemma-3 4B. Our selection included both closed-source proprietary models (GPT-5, GPT-4o) and open-source models (Llama-3, Mistral, Phi-4, Gemma-3); we initially attempted to use Google's Gemini series, but its content safety filters prevented testing, so we substituted Google's Gemma-3 as an alternative from the same provider. Open-source models were selected to have comparable parameter sizes (ranging from 4B to 14B parameters) to enable fair comparison. These models were chosen for their prominence in the field and alignment with our data collection timeline. We accessed closed-source models via official APIs and open-source models through \citet{ollamaOllama2025a}.

\subsection{Climate Change Opinion Measures}\label{subsec:measures}

We treated each question as an individual item rather than combining related questions into composite scales. In a composite scale approach, we would have averaged responses to multiple related questions, e.g., three questions about worry into a single "worry index." Instead, we analyze each question separately. This preserves the granular measurement necessary to detect how biases vary across intersecting identities: we can identify whether Black women are systematically biased on one question (belief in climate causes) while showing accurate responses on another (support for clean energy policies), patterns that would be lost if we collapsed related questions into single indices. The complete list of all 20 survey questions with response options is provided in Appendix~\ref{app:study2}.

Questions employed varied response formats. Some used binary yes/no formats, while others utilized Likert-type scales. For example, the worry question asked: "How worried are you about global warming?" with response options: 1 = Not at all worried, 2 = Not very worried, 3 = Somewhat worried, 4 = Very worried. Another question asked: "When do you think global warming will start to harm people in the United States?" with options: 1 = Never, 2 = In 100 years, 3 = In 50 years, 4 = In 25 years, 5 = In 10 years, 6 = They are being harmed right now.

The response options shown to the models were generated from the survey codebook, so any category the instrument coded was passed through to the prompt. Nine of the twenty items carry a "don't know" or "not sure" category in that codebook and therefore presented it to the models; the remaining eleven do not. The split follows the kind of question being asked. The instrument offers a don't-know wherever an item asks what is true or what will happen: whether global warming is happening, what causes it, what scientists think, and the five items asking how much harm it will do. It withholds one wherever an item asks what the respondent wants, feels, or does, which covers worry, discussion frequency, and the eight policy items. The timing item is the instructive exception, since it asks for a prediction but offers "Never" as a substantive answer, leaving a skeptic somewhere to land without an abstention category.

Where the codebook places that category determines what the models could do with it. On the five harm items and the media-exposure item the codebook assigns "don't know" the value 0, which our system prompt excluded by restricting responses to the range 1 through 6, so the models could not select it; their zero abstention rate on those six items is an artifact of the instruction rather than a behavioral result, and we draw no inference from it. On the three belief items the category sits inside the numbering (2 for whether global warming is happening, 1 for its cause and for scientific consensus), so the models could select it and a comparison with human respondents is meaningful. Of those three, the cause item is unusable in practice: no human respondent selected the category, leaving no base rate to compare against. Appendix~\ref{app:matched} reports the remaining two.

The CCAM survey instrument has been used extensively in climate change research, though detailed psychometric properties for individual items were not publicly available from the survey administrators. We relied on the established validity of these measures demonstrated through their widespread use in peer-reviewed publications \citep{ballewClimateChangeAmerican2019, bergquistEnergyPolicyPublic2020, cutlerGlobalWarmingAffecting2020, hurleyHowPeopleChange2025, marlonChangeUSStatelevel2022a, stewartMappingRacialEthnic2024}.

\subsection{Procedure}

Data collection occurred from September through November 2025, including multiple sensitivity analyses (see Appendix~\ref{app:senst}). For each of the 978 demographic profiles, we prompted each LLM to assume the identity of an individual with specific sociodemographic characteristics matching those of actual survey respondents. 

The demographic persona prompt followed this template:
\begin{quote}
\begin{Verbatim}
You are a [age] year old [race] [gender] with a [education level] with a household income of [income category]. You consider yourself [ideology] and identify as [party identity (including partisan leaners)]. You live in the [region]. You are [are not] religious.
\end{Verbatim}
\end{quote}

For example: "You are a 40 year old White, non-Hispanic male with a bachelor's degree or higher with a household income of \$100,000 or more. You consider yourself somewhat conservative and identify as Republican. You live in the West. You are religious."

We queried each LLM separately for each of the 20 climate questions, repeating the demographic persona prompt before each question to maintain consistent role-playing. Each of the 20 questions was submitted as an independent API call with no conversational memory between queries. The demographic persona prompt was included in every call, ensuring that each question was answered without any context from prior questions. Since no information carried across calls, the order in which questions were queried should not affect results. Questions were presented in the same order across all profiles and models. We used a structured response format requiring LLMs to select only the numeric option corresponding to their answer. We presented response options in the original survey order rather than randomizing them to maintain symmetry with human respondents who received fixed-order options; while LLM responses can be sensitive to option ordering \citep{dominguez-olmedoQuestioningSurveyResponses2024}, our goal is to assess LLM-human agreement under matched conditions rather than to isolate prompt sensitivity effects.

LLMs receive two types of instructions: a system prompt, which sets persistent behavioral constraints for the session, and user prompts containing the actual queries. We used the following system prompt:

\begin{quote}
\begin{Verbatim}
You are taking a survey. You must respond with only a single number (1, 2, 3, 4, 5, or 6) that corresponds to your choice. Do not provide explanations, reasoning, or additional text.
\end{Verbatim}
\end{quote}

The demographic persona and survey questions were provided as user prompts, with the response format instruction appearing in both prompts to reinforce compliance with the structured output requirement. The user prompt template was:

\begin{quote}
\begin{Verbatim}
[Demographic persona prompt]

Based on your background and perspective, please answer the following question:

[Question text]

Please choose from these options:
[Numbered response options]

Respond with only the number corresponding to your choice (e.g., "1", "2", "3", etc.).
\end{Verbatim}
\end{quote}

The closing instruction gave generic example numbers rather than the valid options for the item at hand, so no item-specific range reached the model from either prompt.

To minimize response variability and capture each model's central tendency for representing demographic perspectives, we set temperature to 0 for every model that permits it. Sampling parameters beyond temperature were applied per provider rather than uniformly. The four models served locally through Ollama received top\_p = 0.1 (restricting sampling to the most probable 10\% of the distribution) and repeat\_penalty = 1.1 (slightly discouraging repetitive outputs). The two OpenAI models received neither: our configuration passed no top\_p, so the API default of 1.0 applied, and repeat\_penalty has no OpenAI equivalent. These settings prioritize reproducibility and model consistency over response diversity, following recommendations to average out variance from decoding randomness \citep{reuelBetterBenchAssessingAI2024, bidermanLessonsTrenchesReproducible2024}.

To validate these choices, we conducted sensitivity analyses for both temperature (0.1, 0.5, 1.0) and top\_p (0.1, 0.5, 0.9) using baseline prompts without demographic personas; neither parameter affected mean responses or variance across any model (see Appendices~\ref{app:temperature} and \ref{app:top_p}). Top\_p and temperature are alternative sampling controls that serve similar functions \citep{CreateModelResponse}. For structured tasks requiring single-token numeric responses, the model's probability distribution is concentrated on one or two tokens, rendering both sampling parameters functionally inert. Appendix~\ref{app:decoding} tests the claim under persona conditioning and finds top\_p exactly inert at temperature 0: holding each model fixed and moving top\_p across its usable range leaves every response unchanged. GPT-5, a chain-of-thought model that only permits temperature = 1 and top\_p = 1, also showed comparable response patterns to other models at their default settings.

Each demographic profile was queried once per LLM for each question under the primary (full demographic) condition, generating 978 profiles $\times$ 6 LLMs $\times$ 20 questions = 117,360 expected responses.

\subsection{Data Quality and Preprocessing}

LLMs generally complied with instructions to provide numeric responses. When models produced verbose responses, we used automated parsing scripts to extract the numeric answer. Script accuracy was validated through manual checking of a random sample of responses. Where a model failed to return a usable response we re-queried once. See Table~\ref{tab:desc_stats_question_model} for full response counts by question and model. 

All six models returned a parseable numeric answer to every query, so no case was excluded for model non-response. A further 633 LLM responses (0.6\% of the analytic set) fell outside the option range of the item they answered, a consequence of the response instructions carrying no item-specific bounds; we excluded these as parse failures rather than treating them as substantive answers.

For the human survey data, we treated blank and refused entries as missing and dropped them at the question level. The CCAM instrument offers a ``don't know'' category on nine of the twenty items, and where it codes that category outside the substantive response scale, as it does on the five harm items and the media-exposure item, we dropped those responses as well. On two items the category falls inside the response ordering: code 2 of three on whether global warming is happening, and code 1 of four on scientific consensus. There we retained the responses at their coded position, so per-question sample sizes range from 872 to 978 across the 20 items (Table~\ref{tab:desc_stats_human}). Appendix~\ref{app:dkrobust} reports the alternative, in which don't-know responses are dropped on every item that offers one.

Six respondents had missing religiosity data; these were assigned the default value of "not religious" in the LLM prompt, and excluding these cases did not change results. We examined response distributions for extreme values. Since extreme responses represent substantively meaningful variation in climate change opinions rather than measurement error, we retained all values in analyses.

\subsection{Sensitivity Analyses}

To assess whether demographic effects operate independently or through interactions, we conducted two sensitivity analyses: single-variable tests and saturation analyses (see Supplementary Information for full results). Due to computational constraints, we randomly sampled 100 respondents from our full sample of 978 and repeated this procedure five times using different random seeds, generating five independent subsamples. Recomputing each effect across subsamples drawn without replacement is the logic of subsampling \citep{politisSubsampling1999}, and pooling the five draws prevents any single random sample from driving the result, a precaution warranted by the run-to-run variability of language-model outputs \citep{bidermanLessonsTrenchesReproducible2024}.

First, we conducted single-variable tests presenting only one demographic characteristic at a time (e.g., "You are a female"), systematically examining each of the seven demographic variables individually. Second, we conducted saturation analyses progressively reducing the number of demographic variables from the full seven-variable persona, removing variables in order of their predictive importance for climate opinions based on prior literature \citep{hornseyMetaanalysesDeterminantsOutcomes2016, dunlapWideningGapRepublican2008}. All sensitivity conditions used identical querying procedures as the main analysis.

These analyses revealed that both gender and race effects reverse direction when tested in isolation compared to the full-persona specification. The gender effect, which is negative (underestimating female responses, i.e., predicting lower pro-climate opinion than humans reported) in the full persona, becomes positive (overestimating female response, predicting higher pro-climate opinion) when gender is the only known attribute. Similarly, the race effect reverses from positive (overestimating Black respondents) to negative (underestimating Black respondents). These reversals support our finding that demographic biases operate through interactions, particularly the gender-by-race interaction, rather than as independent main effects. The saturation analysis further shows that the gender effect remains stable across persona configurations with two or more variables, reversing only when completely isolated. Full results are reported in  Appendix~\ref{app:senst} .

\subsection{Data Diagnostics and Analytical Strategy}

Prior to data collection, we established our data processing and analysis plan. We planned to use fixed effects regression models with clustering at the respondent level to account for the nested structure of our data. We used $\alpha = .05$ as the threshold for statistical significance.

Our primary outcome was the response gap between LLM-generated and human responses. To enable comparison across questions with different response scales, we normalized all responses to a 0--1 scale using theoretical minimum and maximum values for each question, defined by the response format (e.g., for a 5-point Likert scale, $r_{\min}=1$, $r_{\max}=5$ and for a 4-point Likert scale, $r_{\min}=1$, $r_{\max}=4$):

\begin{equation}
r^{\text{scaled}} = \frac{r - r_{\min}}{r_{\max} - r_{\min}}
\end{equation}

This ensured consistent scaling across questions and preserved interpretability of the full response range. The response gap ($g = r^{\text{scaled}}_{\text{LLM}} - r^{\text{scaled}}_{\text{human}}$) captures both direction and magnitude of misrepresentation: a positive gap (overestimation) means the LLM assumes the respondent is more pro-climate than they actually are; a negative gap (underestimation) means the opposite.

We estimated fixed effects regression models to examine demographic patterns in LLM representation accuracy:

\begin{equation}
\begin{split}
g_{ijkm} =\ & \beta_1 \text{Gender}_i + \beta_2 \text{Age}_i + \beta_3 \text{Race}_i + \beta_4 \text{Ideology}_i \\
& + \beta_5 \text{Education}_i + \beta_6 \text{Income}_i + \beta_7 \text{Region}_i + \\
& \beta_8 \text{Religiosity}_i + \alpha_j + \gamma_k + \epsilon_{ijkm}
\end{split}
\end{equation}

where $i$ indexes demographic profiles, $j$ indexes questions, $k$ indexes LLM models, and $m$ indexes observations. The $\beta$ coefficients are fixed parameters estimated across all observations; demographic variables carry subscript $i$ because they vary only by profile. The model included fixed effects for questions ($\alpha_j$) and LLM models ($\gamma_k$) to account for systematic differences across questions and models. Standard errors were clustered at the individual respondent level to account for repeated observations within demographic profiles across questions and models.

The design makes these rows dependent in a specific way. Pairing 978 respondents with six models and twenty items gives 117,360 cells, of which 113,373 remain after exclusions, so each respondent contributes up to 120 observations. A one-way variance decomposition of the pooled model's residuals locates that dependence: 27\% of residual variance lies between respondents, while the share lying between models is zero to three decimal places once model fixed effects are absorbed. Residuals for the same respondent correlate at .785 on average across the fifteen model pairs, ranging from .549 to .946. Much of that correlation is structural rather than behavioral, since every model's gap for a respondent is measured against the same human answers, so a respondent whose answers depart from their demographic baseline produces a similar residual under every model.

Our specification answers that structure rather than ignoring it. Standard errors cluster on the respondent, which permits arbitrary correlation across the models and items belonging to one person and therefore treats the 978 respondents, not the 113,373 rows, as the independent units. We do not cluster on the model, since the fixed effects already absorb between-model variation and six clusters would be too few for cluster-robust inference; the six models are instead compared directly, one estimate each, in Figure~\ref{fig:coef_divergence}. We do not add respondent fixed effects, which would absorb the demographic variables the analysis is about, and we prefer cluster-robust errors to a random-effects specification, which would require the respondent effect to be uncorrelated with those same demographic regressors.

We examined model assumptions through residual diagnostics. Residual plots showed no obvious heteroskedasticity. Quantile-quantile plots indicated approximately normal residual distributions. Variance inflation factors were all close to 1, indicating no concerning multicollinearity among covariates.

To examine intersectional patterns, we estimated interaction models testing whether representation gaps varied across combinations of demographic characteristics. We also conducted stratified analyses estimating separate models within racial/ethnic subgroups to examine how other demographic characteristics (gender, ideology, age, education, income, region, religiosity) related to representation gaps within each racial/ethnic group.

For demographic variables, we used the following coding: gender (male, female), age (continuous in years), race (White non-Hispanic, Black non-Hispanic, Hispanic, Other non-Hispanic), ideology (5-level categories: very liberal to very conservative) rather than a continuous scale to capture non-linear effects, education (less than high school, some college, bachelor's degree or higher; coded ordinally), income (less than \$50,000, \$50,000--\$99,999, \$100,000 or more), region (Northeast, Midwest, South, West), and religiosity (not religious, religious; coded 0/1). While prompts included party identity to create realistic personas, we modeled ideology rather than party because ideology captures the directional variation in climate opinions and is standard in climate opinion research \citep{goebbertWeatherClimateWorldviews2012, czarnekRightwingIdeologyReduces2021}; party identity and ideology are moderately associated (Cramer's V = $0.41$), so including both party identity and ideology in the model would introduce multicollinearity concerns.

While this study was not preregistered, the research question, design, and analytical strategy were established prior to data collection. The intersectional patterns reported here emerged consistently across multiple models and demographic configurations, and sensitivity analyses (single-variable tests, saturation analyses) suggested that the overall finding of compression and race-specific gender effects was consistent across analytical specifications.

LLM data collection was implemented in Python version 3.11.0, and all statistical analyses and their visualizations were conducted in R version 4.5.1 using the \texttt{fixest} package for fixed effects estimation.

The analysis supports three claims, and they are not interchangeable. Each is defined at a different unit, estimated by a different specification, and limited in a different way, so that evidence for one does not carry to another.

\textit{Prediction error.} The unit is one respondent, one model, one item, giving 113,373 observations. We regress the scaled gap on the seven persona variables, with question and model as fixed effects and standard errors clustered on the respondent. We read the direction of the gap rather than its level, since conditioning on demographics alone leaves some gap expected by design. Two people with the same demographic profile can answer differently, so no model could place every respondent exactly.

\textit{Variance compression.} The unit is the item, pooling respondents and models within each of the twenty questions. We take one minus the mean across items of model variance divided by human variance. The quantity describes the spread of a distribution rather than any single prediction, and the measured value depends on the decoder (Appendix~\ref{app:decoding}). A synthetic sample can match a population's average opinion and still contain almost none of its disagreement.

\textit{Intersectional reversal.} The unit is again one respondent, one model, one item, but estimated within a racial group. We fit the same specification with a gender by ideology interaction separately in each of the four groups. We claim the sign of the difference rather than its size, since the smallest cells cannot support a point estimate (Appendix~\ref{app:smallcell}). The evidence supports saying that the error runs the other way for Black respondents, not how far it runs.

\section{Results}\label{sec:results}

Throughout this section, positive gaps are overestimates and negative gaps underestimates.

\subsection{Overall Systematic Compression of Predicted Climate Opinions}\label{subsec:compression}

\begin{figure}
    \centering
    \includegraphics[width=.9\columnwidth]{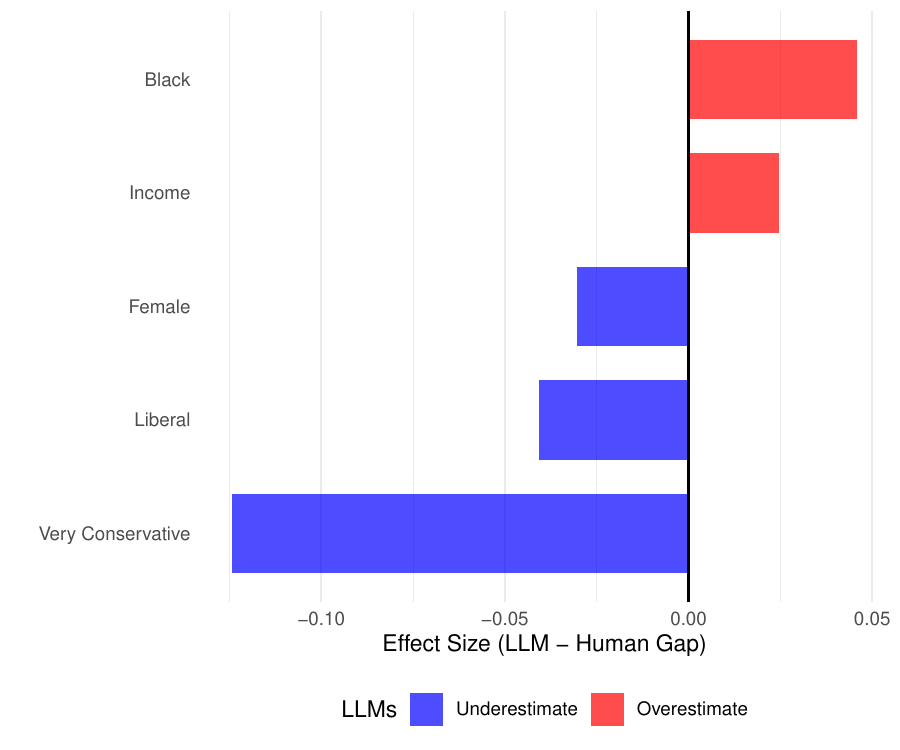}
    \caption{Statistically Significant Demographic Effects on Prediction Gaps. Bars show coefficients where $p < .05$ from regression analysis of LLM-human response differences. Positive values (red) indicate LLM overestimation in the pro-climate direction; negative values (blue) indicate underestimation.}
    \label{fig:simple_bar_plot}    
\end{figure}

We first examined whether LLMs accurately predict climate opinions across demographic groups by regressing LLM—human response gaps on demographic characteristics with fixed effects for questions and models (Figure~\ref{fig:simple_bar_plot}; Table~\ref{tab:main}). Five demographic variables showed statistically significant prediction differences ($p < .05$). LLMs overestimated climate concern for Black respondents ($\beta = 0.046$) and higher-income individuals ($\beta = 0.025$), while underestimating climate concern for females ($\beta = -0.030$), liberals ($\beta = -0.041$), and especially very conservative respondents ($\beta = -0.124$). Age showed a marginal effect in the same direction ($\beta = -0.0006$ per year, $p < .1$). 

\begin{figure}
    \centering
    \includegraphics[width=0.9\linewidth]{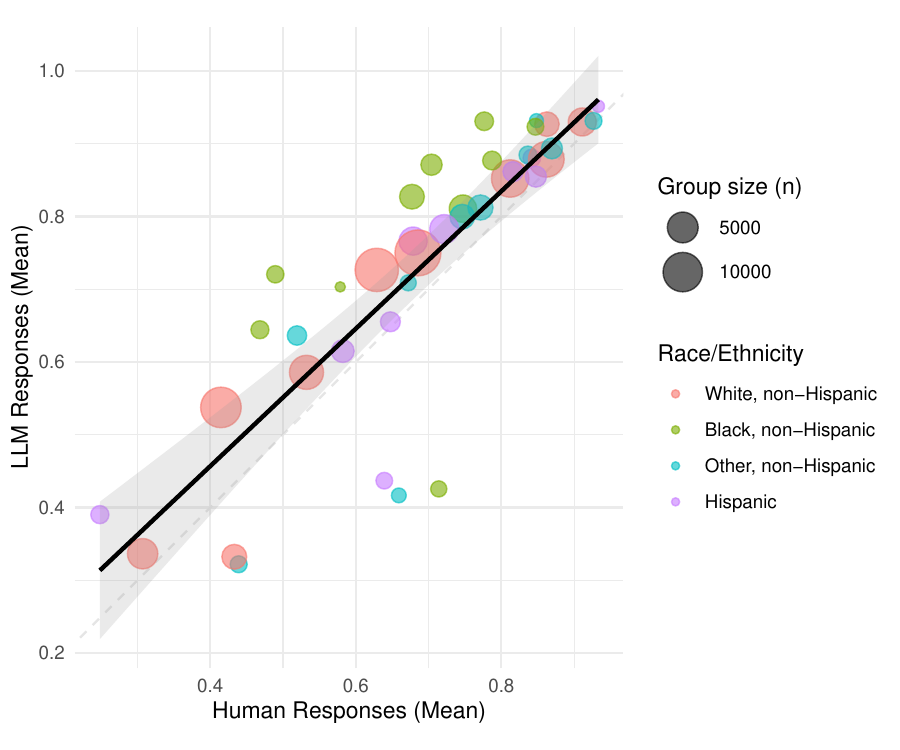}
    \caption{\titlecap{Compression of predicted responses toward population mean.}\label{fig:funnel_plot} Each point represents mean responses for a demographic subgroup (n = group size), with race/ethnicity indicated by color. The diagonal line represents perfect prediction accuracy; the gray band shows 95\% confidence interval. Points above the line indicate overestimation in the pro-climate direction; below indicates underestimation. Note that prediction mismatches vary systematically by race, with Black subgroups (green) consistently above the line, a pattern explored in subsequent sections.}
\end{figure}

Beyond directional biases, LLMs exhibited a pattern of variance compression (Figure~\ref{fig:funnel_plot}). When comparing mean predicted responses to actual human responses across demographic subgroups, LLM predictions clustered more tightly around the population mean than human responses. Groups with lower actual climate concern were overestimated (points above the diagonal), while groups with higher actual concern were underestimated (points below the diagonal). This compression narrows the predicted range of American climate opinions. Prediction mismatches also varied systematically by race: Black subgroups (green) appear mostly above the diagonal line, indicating overestimation across most subgroups, while White, Hispanic, and Other subgroups cluster closer to the line with mismatches in both directions.

Moreover, across all 20 climate opinion questions, LLM responses showed on average 29.4\% less variance than human responses (mean variance ratio = 0.71) (Table~\ref{tab:variance}). This compression pulls predictions toward the population mean, understating the diversity of American climate opinions. The same pattern appears in how rarely the models abstain: on the two items where humans and models chose from identical option sets, the models selected ``don't know'' less often than respondents did, markedly so on scientific consensus (Appendix~\ref{app:matched}). These aggregate patterns, however, obscure important heterogeneity. The overall gender effect (female $\beta = -0.030$) masks substantial variation across racial groups, which we examine in the following section.

\subsection{Gender Mismatches Vary by Race in LLM Predictions}\label{result:racialized-gender}

\begin{figure}
    \centering
    \includegraphics[width=0.95\linewidth]{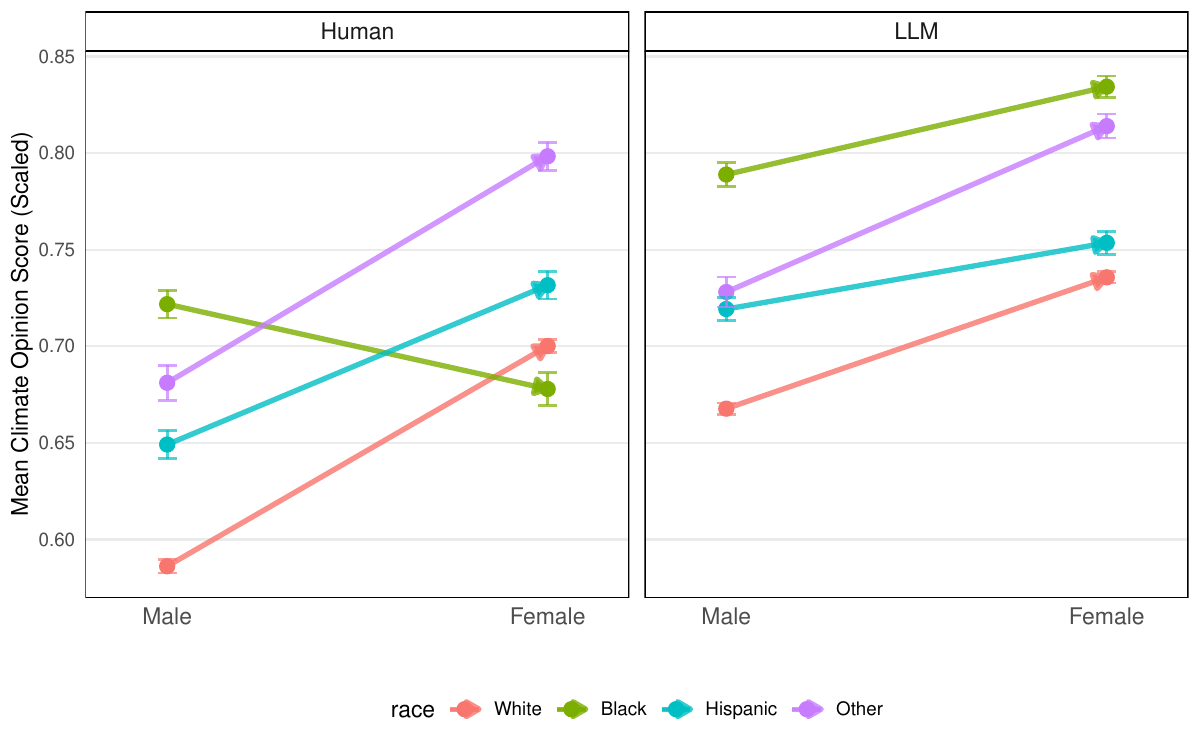}
     \caption{\titlecap{Human responses versus LLM predictions by gender and race.} Mean scaled responses by gender within each racial group, comparing actual human responses to LLM predictions. Error bars represent standard errors. Results suggest LLMs may apply a uniform gender pattern that does not fully capture race-specific differences observed in human data.}
    \label{fig:gender_race}
\end{figure}

 The overall gender effect ($\beta=-0.030$) (Table~\ref{tab:main}) masks a notable interaction: LLMs apply uniform gender patterns that diverge from race-specific patterns. Figure~\ref{fig:gender_race} and Table~\ref{tab:gender_by_race}~(Appendix~\ref{app:descriptive}) compare actual human responses to LLM predictions. In reality, gender patterns differ by race: White, Hispanic, and Other females report higher climate concern than males of the same race (scaled differences of $+0.113$, $+0.084$, and $+0.117$), while Black females report slightly lower concern than Black males ($-0.042$). LLM predictions show a flattened gender pattern, with females predicted higher than males in all four racial groups ($+0.034$ to $+0.086$)~(Table~\ref{tab:gender_by_race}). The assumption therefore matches the human pattern everywhere except among Black respondents, where it reverses. In the terms set out earlier, the models recognize \emph{Black} and they recognize \emph{woman}, but they do not represent \emph{Black woman} as a distinct position. 

This uniform assumption is associated with opposite prediction mismatches (Table~\ref{tab:main}). For White respondents, LLMs underestimate female climate opinions relative to males ($\beta=-0.045$, $p<.01$). For Black respondents, the gap reverses: LLMs overestimate female opinions relative to males ($\beta=+0.066$, $p<.05$), predicting females are more pro-climate than males, when the opposite is true. Hispanic respondents show underestimation similar to White respondents ($\beta=-0.045$, n.s.).  Other respondents show no statistically significant gender gap ($\beta=-0.041$, n.s.). 

\begin{figure}
    \centering
    \includegraphics[width=0.95\columnwidth]{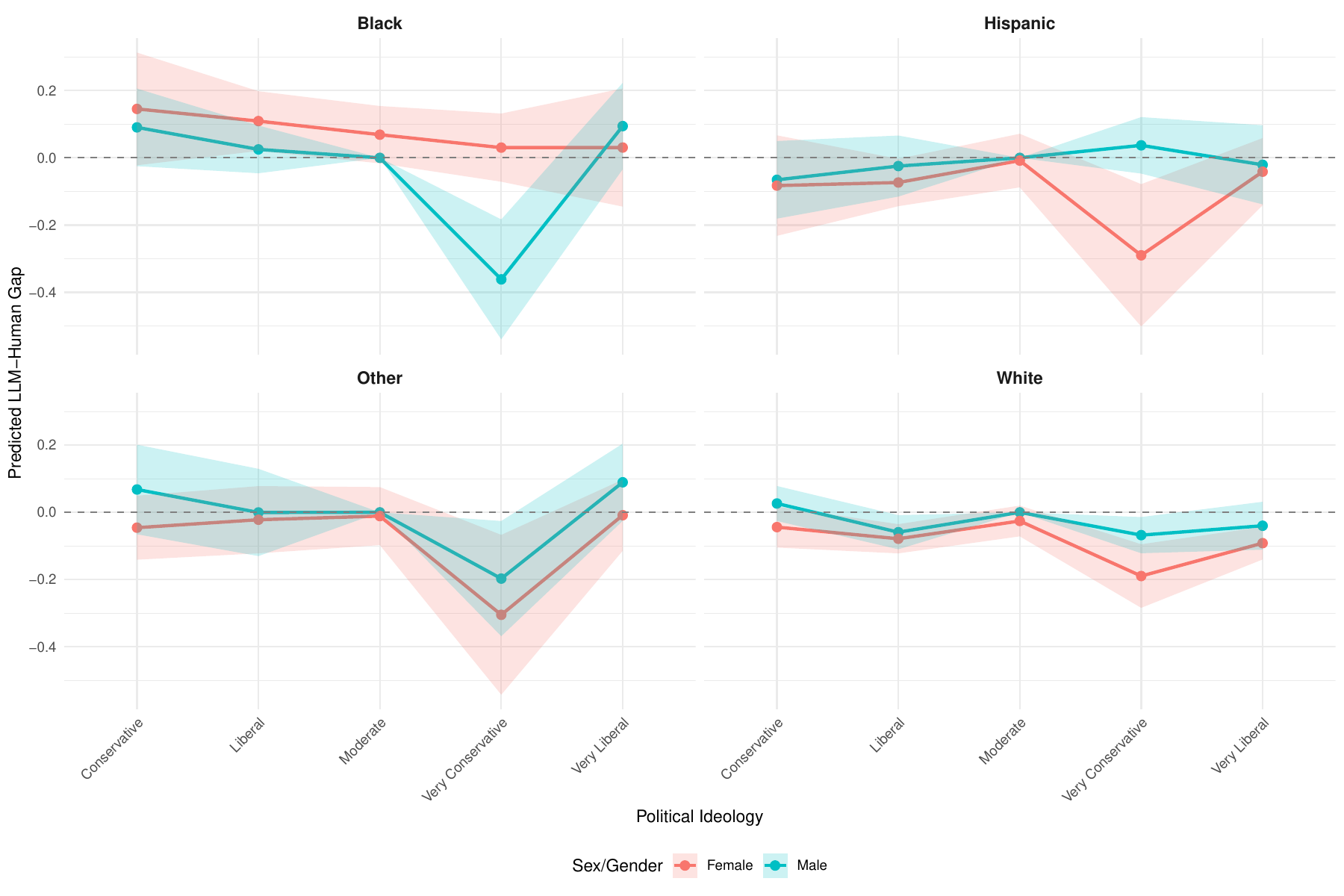}
    \caption{\titlecap{Predicted LLM--human gap by political ideology, gender, and race}. Positive values indicate overestimation in the pro-climate direction; negative values indicate underestimation. Predictions are derived from race-stratified regression models with Gender~$\times$~Ideology interactions, controlling for age, education, income, region, and religiosity (Table~\ref{tab:tab_racial_subset}).  Black and Hispanic respondents show opposite gender patterns at the very conservative extreme (Female~$\times$~Very Conservative: $\beta=+0.323$, $p<.01$ for Black; $\beta=-0.319$, $p<.01$ for Hispanic), both surviving correction for multiple comparisons across the four race-stratified tests, though the estimates rest on small subgroups (Table~\ref{tab:cell_sizes}). White respondents show a negative interaction that does not survive that correction ($\beta=-0.096$, $p<.1$), and Other respondents show no statistically significant gender differences. }
    \label{fig:marginal_race}
\end{figure}

  Figure~\ref{fig:marginal_race} shows these patterns vary further across the political spectrum. Among White respondents, gender differences in prediction mismatches are minimal across ideologies; the negative interaction at the very conservative extreme ($\beta=-0.096$, $p<.1$) does not survive correction for multiple comparisons (Appendix~\ref{app:smallcell}). Among Black and Hispanic respondents, the gender gaps are largest at the very conservative extreme and run in opposite directions ($\beta=+0.323$, $p<.01$ for Black; $\beta=-0.319$, $p<.01$ for Hispanic), and both survive that correction. Both estimates rest on small cells at that extreme (Table~\ref{tab:cell_sizes}, Appendix~\ref{app:descriptive}), and the next subsection tests how much the opposition in sign depends on them.  

These contrasting patterns suggest that LLM predictions do not gravitate toward consistent patterns across racial groups. Different intersectional groups, defined by combinations of race, gender, and ideology, are misrepresented in different directions. A single-variable demographic audit examining only overall gender or race effects would likely miss these patterns.

Sensitivity analyses point to this reflecting an interaction effect rather than an artifact of isolated demographic influences. When LLMs receive only gender information (no race), the gender effect reverses to positive ($+0.135$); when race is also included, the effect becomes negative ($-0.030$). Single-variable tests are unlikely to detect race-specific gender patterns (Table~\ref{tab:sensitivity} and Figure~\ref{fig:sensitivity}).

\begin{table}  
\caption{\label{tab:main}Demographic Effects on LLM--Human Prediction Gaps}
\begin{adjustbox}{width=\columnwidth, center}
\begin{tabular}{lccccc}       \tabularnewline \midrule \midrule                                           & LLM-Human(Overall) & White          & Black          & Hispanic     & Other \\          Model:                              & (1)                & (2)            & (3)            & (4)          & (5)\\         \midrule       \emph{Variables}\\       Female                              & -0.030$^{***}$     & -0.045$^{***}$ & 0.066$^{**}$   & -0.045       & -0.041\\                                              & (0.011)            & (0.014)        & (0.031)        & (0.031)      & (0.033)\\          Age                                 & -0.0006$^{*}$      & -0.0008$^{**}$ & 0.0003         & -0.0006      & $5.58\times 10^{-5}$\\                                               & (0.0003)           & (0.0004)       & (0.0008)       & (0.001)      & (0.0008)\\          Race/Ethnicity: Black               & 0.046$^{**}$       &                &                &              &   \\                                              & (0.018)            &                &                &              &   \\          Race/Ethnicity: Other, non-Hispanic & -0.033$^{*}$       &                &                &              &   \\                                              & (0.017)            &                &                &              &   \\          Race/Ethnicity: Hispanic            & -0.015             &                &                &              &   \\                                              & (0.017)            &                &                &              &   \\          Ideology: Very Liberal              & -0.031$^{*}$       & -0.053$^{**}$  & 0.053          & -0.037       & 0.033\\                                              & (0.017)            & (0.021)        & (0.056)        & (0.054)      & (0.040)\\          Ideology: Liberal                   & -0.041$^{***}$     & -0.057$^{***}$ & 0.039          & -0.043       & -0.003\\                                              & (0.013)            & (0.016)        & (0.033)        & (0.032)      & (0.037)\\          Ideology: Conservative              & 0.006              & 0.006          & 0.084$^{*}$    & -0.074       & 0.030\\                                              & (0.016)            & (0.019)        & (0.050)        & (0.048)      & (0.048)\\          Ideology: Very Conservative         & -0.124$^{***}$     & -0.105$^{***}$ & -0.314$^{***}$ & -0.095       & -0.233$^{***}$\\                                              & (0.023)            & (0.025)        & (0.088)        & (0.064)      & (0.075)\\          Education                           & 0.007              & 0.007          & 0.004          & 0.006        & 0.003\\                                              & (0.005)            & (0.006)        & (0.013)        & (0.013)      & (0.014)\\          Income                              & 0.025$^{***}$      & 0.027$^{***}$  & 0.009          & 0.042$^{*}$  & 0.023\\                                              & (0.008)            & (0.009)        & (0.020)        & (0.023)      & (0.025)\\          Region: Northeast                   & -0.006             & -0.012         & 0.065          & 0.021        & -0.038\\                                              & (0.016)            & (0.020)        & (0.041)        & (0.043)      & (0.041)\\          Region: Midwest                     & -0.008             & -0.018         & 0.070          & -0.003       & 0.058\\                                              & (0.016)            & (0.019)        & (0.051)        & (0.061)      & (0.046)\\          Region: South                       & -0.003             & -0.014         & 0.053          & 0.011        & -0.0009\\                                              & (0.014)            & (0.018)        & (0.041)        & (0.034)      & (0.039)\\          Religious                           & 0.027$^{*}$        & 0.010          & 0.088$^{**}$   & 0.079$^{**}$ & 0.081$^{**}$\\                                              & (0.015)            & (0.019)        & (0.043)        & (0.035)      & (0.035)\\          \midrule       \emph{Fixed-effects}\\       Questions                           & Yes                & Yes            & Yes            & Yes          & Yes\\         LLMs                                & Yes                & Yes            & Yes            & Yes          & Yes\\         \midrule       \emph{Fit statistics}\\       Observations                        & 113,373            & 75,224         & 11,888         & 15,742       & 10,519\\         R$^2$                               & 0.14068            & 0.14072        & 0.22006        & 0.16745      & 0.17978\\         Within R$^2$                        & 0.02570            & 0.02704        & 0.08673        & 0.02993      & 0.06400\\         \midrule \midrule       \multicolumn{6}{l}{\emph{Clustered (case\_id) standard-errors in parentheses}}\\       \multicolumn{6}{l}{\emph{Signif. Codes: ***: 0.01, **: 0.05, *: 0.1}}\\    \end{tabular}
\end{adjustbox}
\smallskip
\noindent\parbox{\columnwidth}{\vspace{0.5em}\small\textit{Note:} Overall and Race-Stratified Models. Reference categories are Male (gender), White (race, overall model only), Moderate (ideology), and West (region). Political ideology is coded as a 5-level factor (very liberal, liberal, moderate, conservative, very conservative) rather than a continuous scale to capture non-linear effects; Standard errors clustered by respondent. Signif. Codes: ***: $p < .01$, **: $p < .05$, *: $p < .1$}
\end{table}

\subsection{Gender Underestimation Across All Six Models}

\begin{figure}
    \centering
    \includegraphics[width=0.9\linewidth]{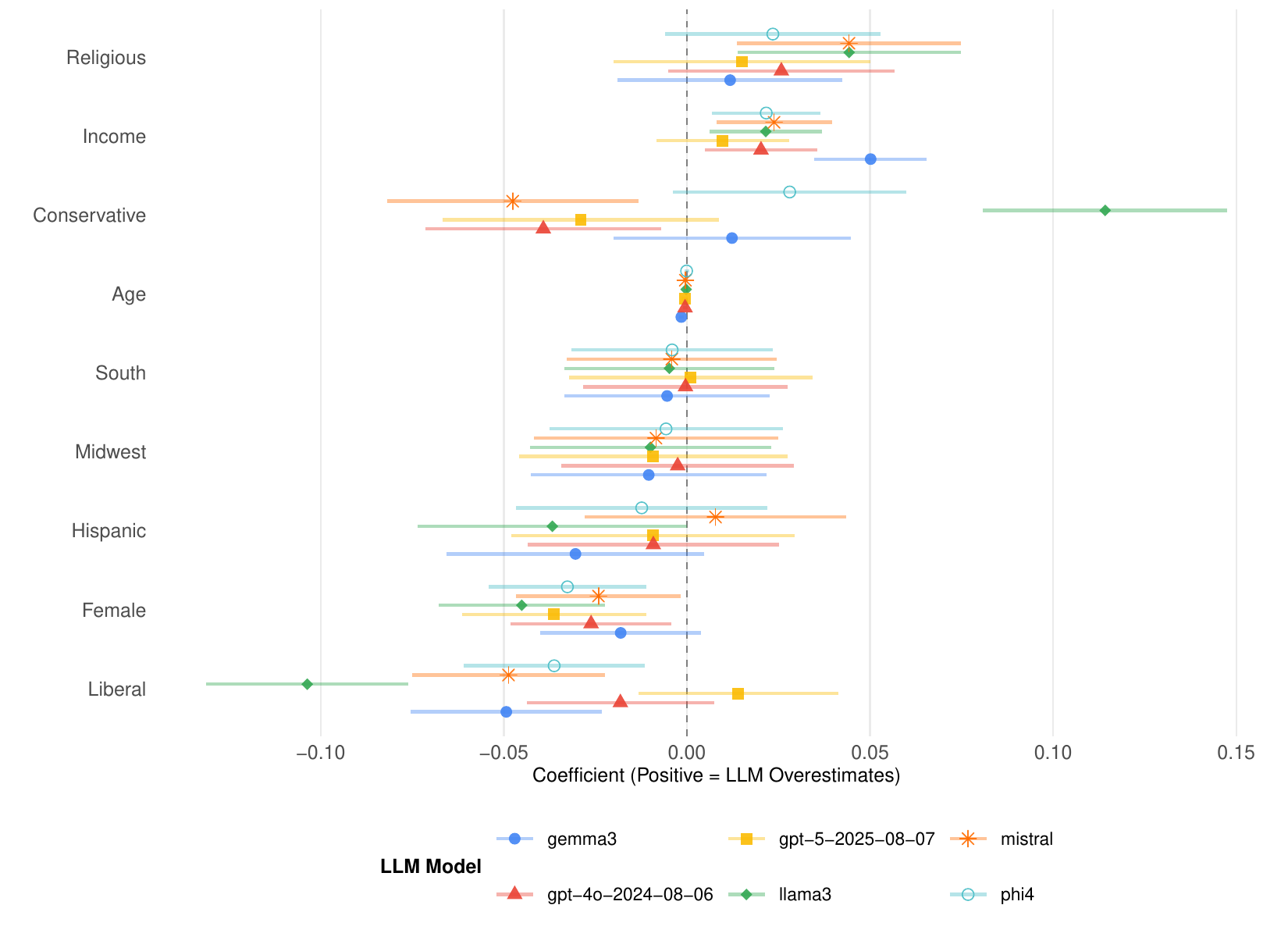}
    \caption{\titlecap{Demographic coefficients across all six models with 95\% confidence intervals}. Points right of zero indicate overestimation in the pro-climate direction; left indicates underestimation. Models diverge on political ideology, with Llama-3 and GPT-4o placing the Conservative coefficient on opposite sides of zero. Female shows negative coefficients across all models.}
    \label{fig:coef_divergence}
\end{figure}

To assess whether demographic prediction patterns reflect idiosyncrasies of specific models or broader tendencies across LLM architectures, we estimated separate regression models for each of the six LLMs (Figure~\ref{fig:coef_divergence}). Models varied markedly in their treatment of political ideology: Llama-3 overestimated conservative climate opinions ($\beta = +0.076$, $p<.01$) while GPT-4o underestimated them ($\beta = -0.027$, $p<.05$). However, all six models showed negative coefficients for gender, underestimating female climate opinions across all architectures tested.

\begin{figure}
    \centering
    \includegraphics[width=0.9\linewidth]{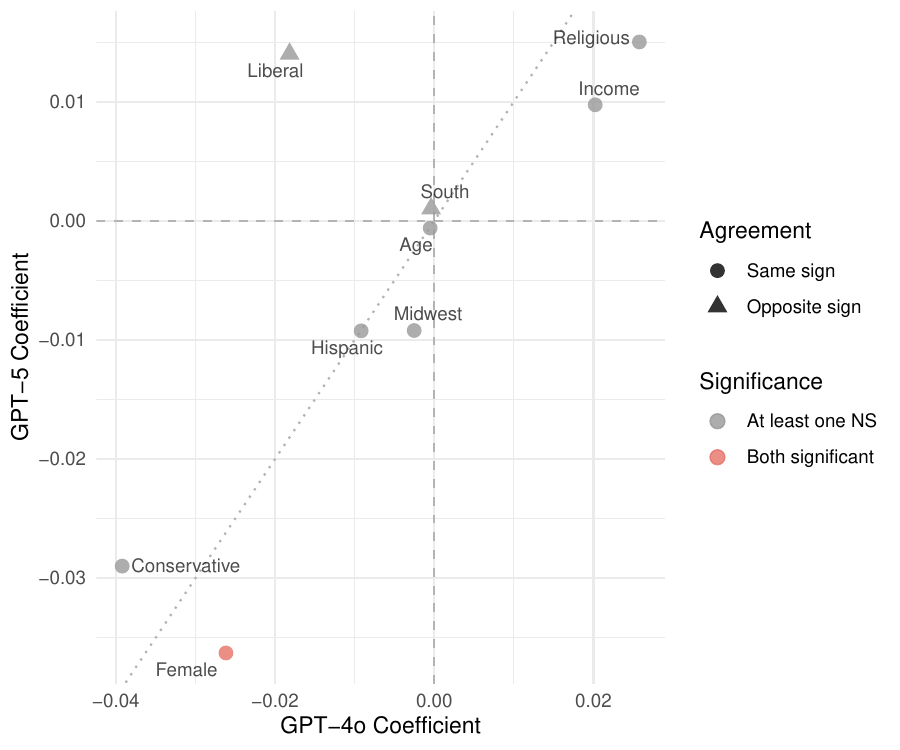}
    \caption{Within-family Comparison (GPT-4o vs. GPT-5). Points along diagonal indicate agreement. Pink indicates both coefficients statistically significant. Female is negative in both models. Most demographics cluster near diagonal, indicating within-family consistency}
    \label{fig:gpt_comparison}
\end{figure}

\begin{figure}
    \centering
    \includegraphics[width=0.9\linewidth]{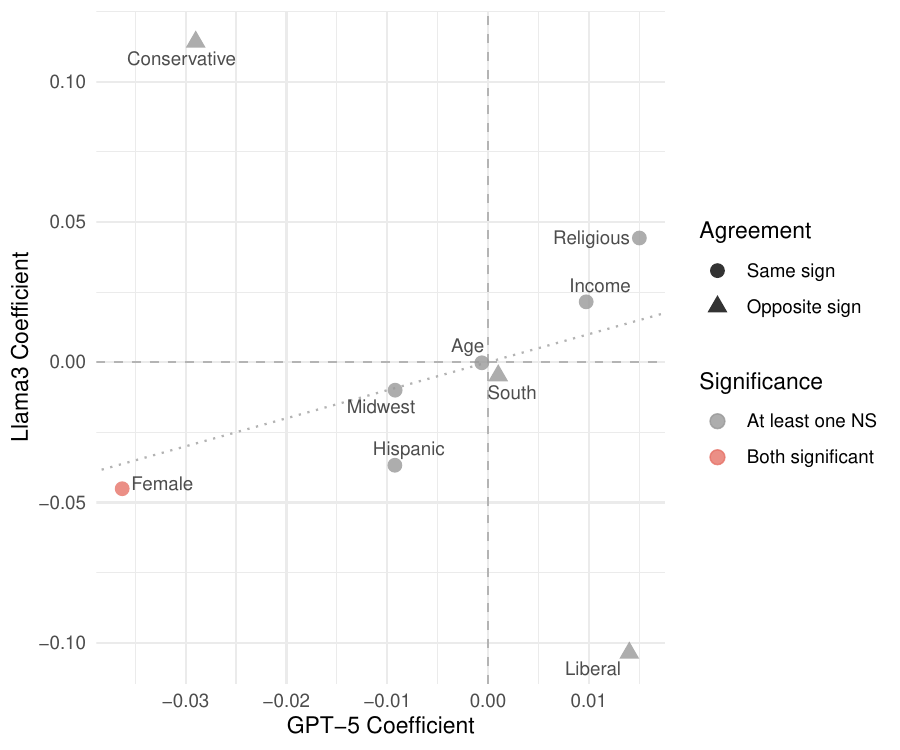}
    \caption{Cross-family Comparison (GPT-5 vs. Llama3). Triangles indicate opposite signs. Conservative and Liberal show opposing coefficients across model families. Female remains negative in both models despite broader architectural disagreement.}
    \label{fig:cross_comparison}
\end{figure}

We further examined model agreement by comparing coefficients across model pairs (Figure~\ref{fig:gpt_comparison}, ~\ref{fig:cross_comparison}). We selected GPT-4o and GPT-5 to test within-family consistency (both OpenAI models), and GPT-5 versus Llama3 to test cross-family consistency (OpenAI versus Meta). Within the GPT family, coefficients clustered along the diagonal, pointing to general agreement. The cross-family comparison revealed systematic divergence: Conservative and Liberal appeared in opposite quadrants with opposing signs, reflecting divergent treatments of political ideology.

Despite this disagreement on ideology, Female remained in the same quadrant for both comparisons: negative in both models, with both effects statistically significant. Gender underestimation is the most consistent demographic pattern across all six models and across both within-family and cross-family comparisons. The models nonetheless differ measurably on other behavioral dimensions, which bounds how far that consistency can be projected (Section~\ref{sec:discussion}). \subsection{Heterogeneous Effects Within Racial Subgroups}

Beyond gender, race also moderates how LLMs represent political ideology. LLMs lean toward underestimating climate opinions at both ends of the ideological spectrum, but asymmetrically. Liberal respondents are moderately underestimated ($\beta=-0.041$, $p<.01$), while very conservative respondents show the largest observed prediction mismatches ($\beta=-0.124$, $p<.01$)(Table~\ref{tab:main}). Conservative and very liberal respondents show no statistically significant effects.

 We focus on very conservatives for two reasons. First, they represent the largest observed prediction mismatch, three times larger than liberals ($\beta=-0.124$ versus $\beta=-0.041$). Second, very conservatives show apparent racial heterogeneity in raw mean LLM--human differences: Black very conservatives lean toward underestimation (gap = $-0.227$), while White very conservatives show minimal mismatch (gap = $-0.016$; Table~\ref{tab:gap_by_race_ideology}, Appendix~\ref{app:descriptive}). This contrast rests on only five Black very conservative respondents (Table~\ref{tab:cell_sizes}), so we treat it as suggestive; it does not appear among liberals, where prediction mismatches are similar across races. 

\begin{figure}
    \centering
    \includegraphics[width=0.9\linewidth]{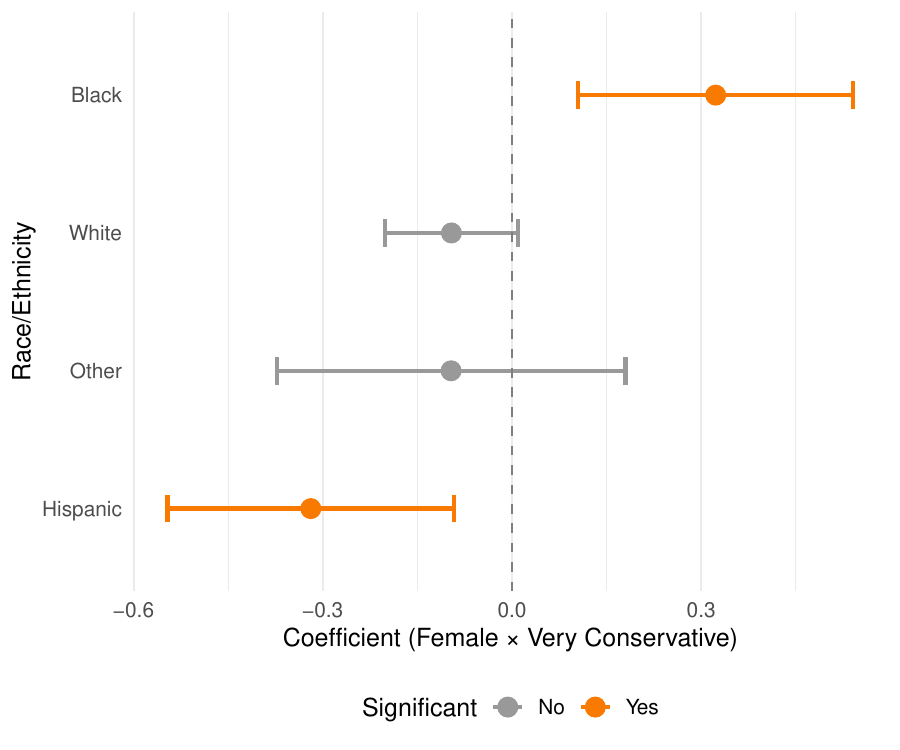}
 \caption{\titlecap{Female $\times$ Very Conservative interaction coefficients by race}. Positive values indicate LLMs predict higher climate opinions for very conservative women relative to very conservative men. Black ($\beta=+0.323$, $p<.01$) and Hispanic ($\beta=-0.319$, $p<.01$) respondents show significant interactions in opposite directions, both surviving Benjamini--Hochberg correction, while White and Other respondents show no gender differences that survive it. These subgroup estimates rest on small cells (Table~\ref{tab:cell_sizes}) and should be read as exploratory.}
    \label{fig:interaction_coef}
\end{figure}

\begin{figure}
    \centering
    \includegraphics[width=0.9\linewidth]{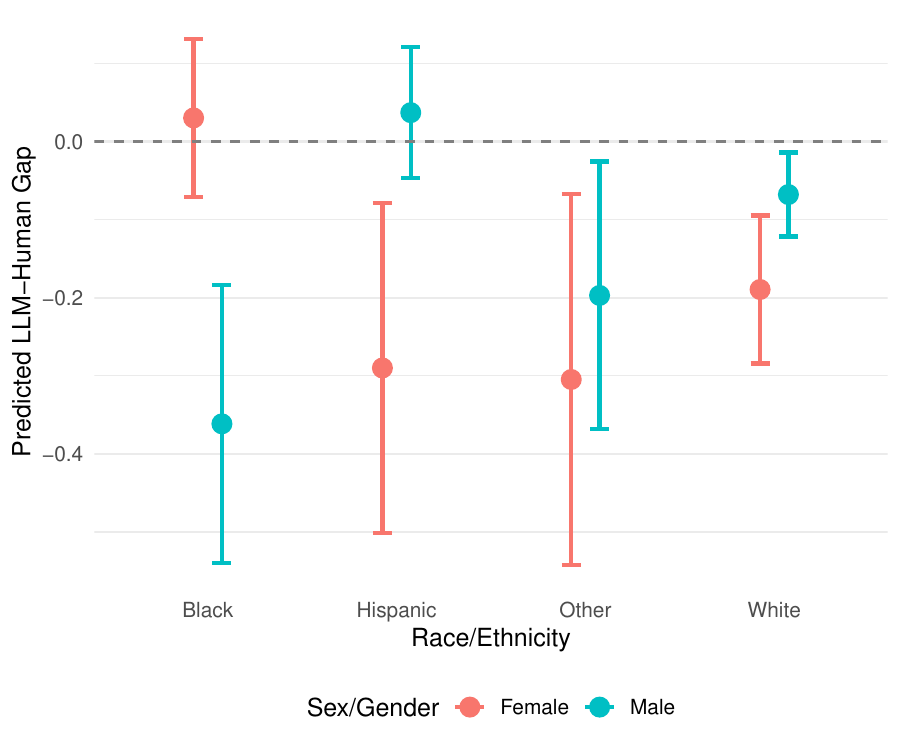}
    \caption{\titlecap{Gender $\times$ Race $\times$ Ideology interaction among very conservative respondents}. Very conservatives show the largest observed prediction mismatches ($\beta = -0.124$), with apparent racial heterogeneity. Black and Hispanic respondents show opposite gender patterns, though based on small subgroups (Table~\ref{tab:cell_sizes}); White respondents show minimal gender differences.}
    \label{fig:marginal_verycons}
\end{figure}

 The race-specific gender pattern in Table~\ref{tab:tab_racial_subset} appears most pronounced among very conservatives (Figures~\ref{fig:interaction_coef} and \ref{fig:marginal_verycons}). Estimating separate models by race, the Female~$\times$~Very~Conservative interaction differs in sign across groups: positive for Black respondents ($\beta=+0.323$, $p<.01$), negative for Hispanic respondents ($\beta=-0.319$, $p<.01$), negative but short of significance for White respondents, and not significant for Other respondents. These very conservative cells are small, with 5 Black respondents (1 female) and 12 Hispanic respondents (Table~\ref{tab:cell_sizes}).

Two checks bound how far the smallest cell drives the result. Collapsing conservatives with very conservatives raises the Black subgroup from 5 respondents to 18 and preserves both signs, though neither interaction remains significant in the wider band; both extreme-band interactions survive Benjamini--Hochberg correction \citep{benjaminiControllingFalseDiscovery1995} across the four race-stratified models (Appendix~\ref{app:smallcell}). We therefore treat the magnitudes as exploratory while reading the difference in sign as the finding.

We caution against reading a statistically nonsignificant interaction as evidence of accurate representation. For Other respondents, wide confidence intervals reflect small subgroups rather than demonstrated parity, and misrepresentation may operate along axes our models do not include. Following \citet{bowlegWhenBlackLesbian2008}, we treat null estimates as uninformative about intersectional misrepresentation rather than as evidence against it. 

 In practical terms, raw mean LLM--human differences by gender echo this pattern, though they rest on very few respondents. Among Black very conservatives, males lean toward underestimation (gap = $-0.288$) and females toward overestimation (gap = $+0.125$); among Hispanic very conservatives, the direction reverses (males $+0.142$, females $-0.202$; Table~\ref{tab:gap_by_race_ideology}). Computed over as few as one to seven respondents per cell (Table~\ref{tab:cell_sizes}), these means are best read as suggestive illustrations rather than reliable estimates. Taken together, the results raise the possibility that LLM prediction mismatches compound at intersectional extremes, a possibility that warrants testing in larger samples. 

\section{Discussion}\label{sec:discussion}

 This study asked how large language models represent demographic and intersectional patterns in U.S. climate change opinions. We found that LLMs appear to flatten the diversity of American climate perspectives. Groups who are less concerned tend to be predicted as more concerned, and vice versa, though Black subgroups are generally overestimated across concern levels. Beyond this overall compression, LLMs appear to apply gender assumptions that diverge from observed patterns in race-specific ways. In human data, gender patterns vary by race: women report higher climate concern than men of the same race in every racial group except one. Among Black Americans the pattern reverses, with men reporting slightly higher concern than women. LLMs predict women as more pro-climate than men in all four groups, an assumption that holds everywhere except for Black Americans. By applying one gender assumption across racial groups, LLMs may therefore misrepresent climate opinions among Black Americans. 

\subsection{Scope and limits of the governance implications}\label{subsec:scope}

 This study documents representational patterns rather than downstream outcomes, and that boundary governs what we claim here. Demonstrating that representational compression produces policy harm would require tracing how AI-generated accounts of public opinion shape particular regulatory decisions, a chain of evidence that current research has not established. Our contribution is the prior step: establishing where compression occurs and for whom, so that such studies become possible. A second boundary follows from our design. Federal agencies currently deploy AI tools to summarize public comments, while researchers use large language models to simulate survey respondents; we tested the latter application, not the former. Inferences about how AI-summarized public input reaches policymakers therefore reach beyond our evidence, and we mark them as conditional below. 

 The risk our evidence speaks to most directly is epistemic, and it falls on research rather than governance. Researchers who substitute silicon samples, model-generated responses standing in for real respondents, for survey respondents build theory on outputs that carry the models' assumptions. A study using our six models to examine gender differences in climate concern would recover a positive gender gap in every racial group, including among Black Americans, whose observed gap runs in the opposite direction (Table~\ref{tab:gender_by_race}, Appendix~\ref{app:descriptive}). Conclusions about how gender, race, and ideology jointly shape climate opinion would then reproduce a demographic assumption rather than a population pattern. An audit conducted on gender alone would not expose the problem: it reports that the models underestimate women's climate concern ($\beta=-0.030$, $p<.01$), while for Black respondents the models in fact overestimate it ($\beta=+0.066$, $p<.05$). The reversal becomes visible only at the intersections, where researchers relying on synthetic respondents are least likely to look. 

 Whether the patterns we document matter for governance depends on a deployment we did not test. If compression of the kind we observe extends to summarization, it could mask intersectional differences, so that a report of broad public support conceals divergence by race, gender, and ideology. Outreach built on the universal assumption that women are more pro-climate might likewise misjudge groups for whom our data suggest the opposite, such as Black very conservative men, while overstating engagement needs for groups whose views are already overstated. Since these subgroup patterns rest on few respondents (Table~\ref{tab:cell_sizes}, Appendix~\ref{app:descriptive}), we offer them as hypotheses about where mismatch might matter rather than as demonstrated outcomes. Establishing whether they matter would require studies of AI-assisted public consultation that do not yet exist.

\subsection{Situating Intersectional AI Mismatch in Climate Research}

Our findings add nuance to prior work on LLM representation of climate opinion. \citet{leeCanLargeLanguage2024} found that LLMs underestimated Black Americans' climate beliefs, an important early finding. Our data run in the opposite direction at the aggregate, where the models overestimate Black respondents' climate concern (Table~\ref{tab:main}); the difference from that study likely reflects newer models and a later survey wave, six models on the 2024 wave against GPT-3.5 and GPT-4 on the 2017 and 2021 waves. Our intersectional lens adds a mismatch a single race estimate hides: real Black women report slightly less climate concern than Black men (a female-minus-male gap of $-0.023$), yet the models predict the reverse ($+0.032$; Table~\ref{tab:gender_by_race}), imposing the ``women are more concerned'' assumption they apply to every other racial group. The mismatch sharpens at the ideological extreme, where very conservative Black men show the largest underestimation while very conservative Black women are overestimated relative to them (Table~\ref{tab:gap_by_race_ideology}). Such within-group variation aligns with broader climate opinion research showing that attitudes vary substantially within racial groups, with common predictors operating differently across groups \citep{benegalRaceEthnicitySupport2022, stewartMappingRacialEthnic2024}. Aggregate "Black opinion" or "female opinion" obscures the intersectional complexity that shapes how people experience and perceive climate change.

 These findings extend recent work documenting LLM representational harms. \citet{wangLargeLanguageModels2025a} showed that LLMs flatten within-group diversity across demographic identities; our results are consistent with this pattern in climate opinion but suggest that compression may not be uniform, following ideological lines, with the largest observed distortions at ideological extremes. \citet{wilsonGenderRaceIntersectional2025} found that intersectional bias in employment contexts most disadvantaged Black males. In climate opinion, the intersectional pattern appears structurally different: at the very conservative extreme the gender gap runs in opposite directions across races, the models overestimating Black women relative to Black men but underestimating Hispanic women relative to Hispanic men, though both estimates rest on small subgroups (Table~\ref{tab:cell_sizes}; Table~\ref{tab:gap_by_race_ideology}, Appendix~\ref{app:descriptive}). Even as tentative signals, they suggest that intersectional mismatch is not a fixed pattern across domains but may reflect the specific structure of opinion in each context, which domain-general audits would likely miss. 

Our intersectional findings are consistent with climate opinion research on this point. Support for climate policies varies not only across racial and ethnic groups but within them, with common predictors like partisanship, education, and gender operating differently by group \citep{benegalRaceEthnicitySupport2022}. Climate belief networks differ across racial groups, and within-group centrality of key attitudes varies by education level \citep{stewartMappingRacialEthnic2024}. This complexity is unlikely to be an LLM artifact. It likely reflects how climate attitudes form at intersections of race, gender, and ideology. The concern is that LLMs tend to flatten this documented heterogeneity into blanket demographic assumptions.

 Environmental justice scholarship has long argued that procedural equity, who participates in climate decision-making and how, is foundational to legitimate and effective climate governance \citep{schlosbergDefiningEnvironmentalJustice2007, paavolaFairAdaptationClimate2006}. Inclusive participation improves both adaptation outcomes and policy legitimacy \citep{cattinoDoesPublicParticipation2021, khatibiCanPublicAwareness2021, newigDoesStakeholderParticipation2023}. A related literature cautions that inclusion can be nominal. Token participation changes little unless institutional structures change with it \citep{arora-jonssonVirtueVulnerabilityDiscourses2011a, perkinsClimateJusticeGender2018}, and planning tools are not neutral: they reproduce existing hierarchies unless power and social difference are made explicit \citep{osborneIntersectionalityKyriarchyFramework2015, hankivskyIntersectionalitybasedPolicyAnalysis2014, amorim-maiaIntersectionalClimateJustice2022}. Our findings bear on the simulation case. When intersectional groups are consistently overestimated or underestimated, a synthetic sample does not reproduce the distribution of views it is taken to represent, and the groups whose views diverge most from the demographic average are the ones least accurately rendered. Whether AI-assisted summarization of public comments carries a comparable risk remains untested. Procedural equity requires not only inviting participation but representing it accurately, and simulated respondents do not yet meet that standard. 

\subsection{Contributions}

This study offers several contributions to understanding how language models misrepresent public opinion when they stand in for survey respondents, which is the setting we tested rather than any deployed governance application (Section~\ref{subsec:scope}).   Our results replicate, in an underexplored domain, the opinion compression that prior work observed in synthetic survey data; the distinct contribution is to show that compression is not uniform but structured by intersecting identities, producing offsetting mismatches that single-variable audits cannot detect.  To our knowledge, it is among the first to examine intersectional misrepresentation in this domain, extending work from healthcare and employment to a policy area where public opinion increasingly shapes decision-making. Our intersectional design, which estimated interactions rather than main effects alone, identified patterns that may be invisible to standard demographic audits. The sensitivity analysis illustrated why this matters: single-variable tests produced conclusions that diverged from full-persona evaluations, suggesting the limits of conventional audit methods. We validated LLM outputs against 978 human survey respondents rather than synthetic benchmarks, quantified prediction gaps in both direction and magnitude, and evaluated six models from different providers to assess generalizability.

A methodological implication follows for silicon-sample work. Compression is a claim about variance, so a decoder that adds randomness inflates the measured variance ratio without improving any individual prediction. Holding personas and items fixed and raising temperature from 0 to 1 lowered measured compression for every model we could test at both settings, by an amount proportional to that model's self-consistency (Appendix~\ref{app:decoding}). Apparent diversity in a synthetic sample is therefore partly a property of the sampler rather than of the opinions represented. Studies reporting variance or diversity in silicon samples should report decoding settings and repeated generations alongside them, since settings alone do not establish stability: GPT-4o disagrees with itself at temperature 0, and GPT-5 cannot be set below 1 at all.

\subsection{Limitations and Future Directions}

 A theoretical limitation precedes the empirical ones. \citet{hoffmannWhereFairnessFails2019} argues that fairness work organized around antidiscrimination categories reproduces the logic it sets out to correct, attending to discrete disadvantages rather than the structures generating them, and \citet{ovalleFactoringMatrixDomination2023} shows how readily such work flattens intersectionality into subgroup metrics. Both criticisms apply to this study. \citet{kongAreIntersectionallyFair2022} argues that the proper object is intersections of oppression rather than of protected attributes, a standard our categorical design does not meet, and \citet{blodgettLanguageTechnologyPower2020} finds that bias research often leaves unstated what makes a behavior harmful. We have named ours as representational, and we claim no more. Intersectional scholarship is not automatically social justice work \citep{collinsIntersectionalityRevisited2016}. Yet \citet{crenshawDemarginalizingIntersectionRace1989} observed that remedies designed from the center exclude those at the margins, and no remedy can begin where the harm has not been located. Auditing is that prior step, and these systems are already in use. 

 Several empirical limitations follow.  While we observe consistent intersectional patterns in LLM misrepresentation, we cannot determine their ultimate causal source. Whether these patterns reflect imbalances in training data composition, inherent model architecture limitations, or emergent properties of large-scale language modeling remains an open question. Second, our sample derives from the Climate Change in the American Mind survey, which is nationally representative by design but may not capture all segments of U.S. public opinion, particularly those underrepresented in online panels. Third, our findings are limited to U.S. climate opinion and may not generalize cross-nationally, where demographic categories carry different social meanings. Finally, LLM outputs are not static: model updates and fine-tuning may alter the patterns we observed. The instability therefore argues for repeated measurement rather than a one-time audit, for any use that depends on these models behaving consistently.

 A further caution concerns the source of the mismatches we document, since several mechanisms could produce them and our design cannot fully separate them. The first is demographic stereotyping, in which the model applies uniform associations to a group, such as assuming women are more pro-climate regardless of race, that do not hold for every subgroup. The second is a limit of demographic-only prompting: because demographic profiles, even with ideology, do not fully determine climate attitudes, a model reasoning perfectly from the information given would still mismatch individuals whose views depart from their demographic baseline. A third possibility concerns our instrument. Our persona prompt presents identity as a concatenation of separate attributes, echoing the additive question wording that \citet{bowlegWhenBlackLesbian2008} shows can make mutually constitutive identities appear separable, and the format may encourage the models to treat gender independently of race. 

 Separating form from content clarifies what our results can support. The additive form of the models' assumptions may well reflect our prompt structure. The content of those assumptions does not: nothing in a persona description specifies that women should be predicted as more pro-climate, nor that the assumption should hold for White, Hispanic, and Other respondents while reversing for Black respondents. The direction of an assumption comes from associations learned during training rather than from the prompt. Nor would an information limit alone produce the pattern we observe, since incomplete information blurs mismatches rather than reversing their sign across groups. The reversal therefore points toward stereotyping, even if the additive form is partly instrument-induced. A direct test would vary both the attitudinal information supplied and the structure of the persona, comparing additive descriptions against intersectionally framed ones to gauge how much mismatch each accounts for.

The temporal gap between our two data sources deserves separate attention. The human responses come from the December 2024 CCAM wave; the model responses were collected between September and November 2025, roughly ten months later. If U.S. climate opinion moved over that interval, part of what we measure as misrepresentation would be real change. The survey's own history bounds how much movement is plausible. Across its 31 waves, fielded at roughly six-month intervals since 2008 \citep{leiserowitzClimateChangeAmerican2020}, the weighted mean of an item changes by 0.023 between adjacent waves on the normalized scale (median 0.018; 485 item-pairs, fewer than a complete panel would give because not every item was fielded in every wave), and the ninetieth percentile of those changes is 0.050. Over the twelve-month intervals available, mean change is also 0.023, indicating that opinion moves between waves without accumulating steadily in one direction over the intervals relevant here; longer-run trends in this series are documented elsewhere \citep{marlonChangeUSStatelevel2022a, eganClimateChangeUS2017}. Movement is not constant across the series: across the final four intervals, those closest in time to our human data, adjacent-wave change averages 0.012, about half the series-wide figure, so the bound matched to our interval is the smaller one. Against a mean prediction gap of 0.060, drift of typical magnitude is not negligible, and we do not claim the mean gap is free of it.

The two findings the paper rests on are nonetheless structurally insulated from drift of that kind. Compression is a claim about variance rather than about means: a population-wide shift moves the distribution without narrowing it, so for drift to manufacture compression the human distribution would have had to widen over those ten months while the models predicted an earlier, narrower one. The intersectional result is a difference of differences, comparing gender gaps within racial groups. Uniform movement in climate opinion cancels in that contrast. For drift to produce the reversal we report, opinion would have had to move differentially across race, gender, and ideology in the specific pattern observed, which is a considerably stronger assumption than that opinion changed. We therefore read the level of the mean gap with caution and the compression and intersectional patterns as robust to plausible temporal change.

Four further features of the design bound what the results can support. Sampling parameters were held at conservative values in the main analysis, with temperature set to 0 for every model that permits it; sensitivity analyses under both baseline and persona-conditioned prompts indicate that neither temperature nor top\_p markedly affects results for structured single-token responses (Appendices~\ref{app:temperature}, \ref{app:top_p}, and \ref{app:decoding}). Six respondents with missing religiosity data were assigned a default value in the LLM prompt; excluding these cases did not alter findings. Finally, while the CCAM microdata was released after the training cutoffs of most models tested, we cannot fully rule out the possibility that earlier survey waves or aggregate reports informed model training. As noted in the analytical strategy, this study was not preregistered; the specific intersectional patterns should therefore be treated as hypothesis-generating, though the two race-specific interactions survive multiple-comparison correction (Appendix~\ref{app:smallcell}).

The convergence on gender warrants a specific caution. Our six models differ not only in size and architecture but in behavior we can measure directly. They differ in reproducibility: the four locally served models return identical responses to repeated queries at temperature 0, while GPT-4o agrees with itself on 91\% of responses under the same setting, a difference of access mode rather than of model design (Appendix~\ref{app:decoding}). They differ in willingness to abstain: on whether global warming is happening, three of the six never selected the don't-know category while Gemma-3 selected it on more than a third of responses (Appendix~\ref{app:matched}). They also differ in how strongly they condition on ideology: the share predicting that climate change will never harm people in the United States ranges from none to 29.9\% across models (Appendix~\ref{app:missingness}). Agreement on the direction of the gender effect therefore emerges against substantial disagreement elsewhere, which makes the shared pattern more striking than a simple count of six would suggest. It does not establish that the pattern extends to models we did not test, to later versions of those we did, or to deployment settings that sample at higher temperatures, where within-model consistency itself falls as low as 43.6\%.

Our design also stands at a distance from the communities whose views it examines. Simulating a respondent is the methodological opposite of building knowledge with that respondent, and participatory approaches surface lived experience in ways a synthetic sample cannot \citep{sureshIntersectionalFeministParticipatory2022, rigonReviewIntersectionalityClimate2025}. Audits of the kind we report identify where representation fails; they do not restore the voices that failure displaces. 

Several directions merit future investigation. Most fundamentally, research linking intersectional mismatch patterns to specific training corpora or fine-tuning procedures could illuminate causal mechanisms, an essential step toward mitigation. Although our follow-up tests with GPT-4o and GPT-5 suggest some consistency within model families, longitudinal tracking across model updates would clarify whether mismatch patterns are stable features or transient artifacts of particular training runs. Beyond climate, applying intersectional audit frameworks to other policy domains where AI mediates public input, such as healthcare, housing, and criminal justice, could reveal whether the patterns we document are domain-specific or reflect broader representational tendencies. Equally important is understanding how practitioners use these tools: do agencies conducting public comment analysis examine intersectional misrepresentation, or do they accept AI-generated summaries uncritically? Field studies of AI deployment in climate governance could bridge the gap between what we observe in controlled settings and what occurs in practice. Finally, replication with other climate opinion datasets and cross-national samples would test whether these findings generalize beyond the U.S. context and our specific survey instrument. Intersectional analysis is also bounded by cell size in any nationally representative sample; designs that oversample at the intersections of race, gender, and ideology, or that pool several survey waves, would allow the magnitude of these reversals to be estimated rather than only their direction.

 We examined how large language models represent U.S. climate opinions across demographic groups by comparing AI-generated estimates to the responses of 978 survey participants. Two patterns emerged: LLMs appear to compress the diversity of climate perspectives toward a population mean, and they appear to apply uniform demographic assumptions that may produce opposite mismatches for different intersectional groups.

The models assume women care more about climate change than men. The assumption understates the difference for White, Hispanic, and Other Americans and runs the wrong way for Black Americans. The inversion is not confined to one corner of the data. Among White, Hispanic, and Other respondents the models' error runs further in the pro-climate direction for men than for women at every ideology level; among Black respondents it runs further for women at four of five, and the contrast is already visible at the moderate level, where the cells are largest. Four tests in Appendix~\ref{app:smallcell} leave the sign standing and none of them supports a magnitude, so the reversal is what we claim and its size is not.

Even so, such reversals may be invisible to audits that test gender, race, or ideology in isolation, because a marginal estimate reports the average mismatch rather than its change of sign. As researchers increasingly use language models to stand in for survey respondents, the central question becomes not whether these systems misrepresent, but whose voices they render legible and whose they obscure. Climate governance, environmental justice scholars have argued for decades, must reflect the complexity of the communities it seeks to serve. The same standard should guide the technologies we rely on to understand them.

\appendix
\onecolumn

\makeatletter
\renewenvironment{table}[1][htbp]{%
  \@float{table}[#1]\centering\sffamily\small
}{%
  \end@float
}
\renewenvironment{figure}[1][htbp]{%
  \@float{figure}[#1]\centering\sffamily\small
}{%
  \end@float
}
\makeatother

\section{Survey Questions and Response Options}\label{app:study2}

{
{\footnotesize
\begin{longtable}{p{0.11\textwidth} p{0.44\textwidth} p{0.39\textwidth}}
\toprule
Item & Question wording & Response options (numeric code) \\
\midrule
\endhead
\texttt{happening} & Do you think that global warming is happening? & 1 = No; 2 = Don't know; 3 = Yes \\
\addlinespace
\texttt{cause\_recoded} & Assuming global warming is happening, do you think it is caused by human or natural factors? & 1 = Don't know; 2 = Other; 3 = Neither, because global warming isn't happening; 4 = Caused mostly by natural changes; 5 = Caused by human activities and natural changes; 6 = Caused mostly by human activities \\
\addlinespace
\texttt{sci\_consensus} & Which comes closest to your own view of what scientists think about global warming? & 1 = Don't know enough to say; 2 = A lot of disagreement among scientists about whether it is happening; 3 = Most scientists think it is not happening; 4 = Most scientists think it is happening \\
\addlinespace
\texttt{worry} & How worried are you about global warming? & 1 = Not at all worried; 2 = Not very worried; 3 = Somewhat worried; 4 = Very worried \\
\addlinespace
\texttt{harm\_personally} & How much do you think global warming will harm you personally? & 0 = Don't know; 1 = Not at all; 2 = Only a little; 3 = A moderate amount; 4 = A great deal \\
\addlinespace
\texttt{harm\_US} & How much do you think global warming will harm people in the United States? & 0 = Don't know; 1 = Not at all; 2 = Only a little; 3 = A moderate amount; 4 = A great deal \\
\addlinespace
\texttt{harm\_dev\_countries} & How much do you think global warming will harm people in developing countries? & 0 = Don't know; 1 = Not at all; 2 = Only a little; 3 = A moderate amount; 4 = A great deal \\
\addlinespace
\texttt{harm\_future\_gen} & How much do you think global warming will harm future generations of people? & 0 = Don't know; 1 = Not at all; 2 = Only a little; 3 = A moderate amount; 4 = A great deal \\
\addlinespace
\texttt{harm\_plants\_animals} & How much do you think global warming will harm plant and animal species? & 0 = Don't know; 1 = Not at all; 2 = Only a little; 3 = A moderate amount; 4 = A great deal \\
\addlinespace
\texttt{when\_harm\_US} & When do you think global warming will start to harm people in the United States? & 1 = Never; 2 = In 100 years; 3 = In 50 years; 4 = In 25 years; 5 = In 10 years; 6 = They are being harmed right now \\
\addlinespace
\texttt{reg\_CO2\_pollutant} & Support/oppose: Regulate carbon dioxide (the primary greenhouse gas) as a pollutant. & 1 = Strongly oppose; 2 = Somewhat oppose; 3 = Somewhat support; 4 = Strongly support \\
\addlinespace
\texttt{fund\_research} & Support/oppose: Fund more research into renewable energy sources, such as solar and wind power. & 1 = Strongly oppose; 2 = Somewhat oppose; 3 = Somewhat support; 4 = Strongly support \\
\addlinespace
\texttt{reduce\_tax} & Support/oppose: Require fossil fuel companies to pay a carbon tax and use the revenue to reduce other taxes by an equal amount. & 1 = Strongly oppose; 2 = Somewhat oppose; 3 = Somewhat support; 4 = Strongly support \\
\addlinespace
\texttt{generate\_renewable} & Support/oppose: Generate renewable energy (solar and wind) on public land in the U.S. & 1 = Strongly oppose; 2 = Somewhat oppose; 3 = Somewhat support; 4 = Strongly support \\
\addlinespace
\texttt{transition\_economy} & Support/oppose: Transition the U.S. economy from fossil fuels to 100\% clean energy by 2050. & 1 = Strongly oppose; 2 = Somewhat oppose; 3 = Somewhat support; 4 = Strongly support \\
\addlinespace
\texttt{priority} & Should global warming be a priority for the president and Congress? & 1 = Low; 2 = Medium; 3 = High; 4 = Very high \\
\addlinespace
\texttt{priority\_cleanenergy} & Should developing clean energy be a priority for the president and Congress? & 1 = Low; 2 = Medium; 3 = High; 4 = Very high \\
\addlinespace
\texttt{teach\_gw} & Support/oppose: Schools should teach children about the causes, consequences, and potential solutions to global warming. & 1 = Strongly oppose; 2 = Somewhat oppose; 3 = Somewhat support; 4 = Strongly support \\
\addlinespace
\texttt{discuss\_GW} & How often do you discuss global warming with your family and friends? & 1 = Never; 2 = Rarely; 3 = Occasionally; 4 = Often \\
\addlinespace
\texttt{hear\_GW\_media} & About how often do you hear about global warming in the media (TV, radio, newspapers, websites, magazines, etc.)? & 0 = Not sure; 1 = Never; 2 = Once a year or less; 3 = Several times a year; 4 = At least once a month; 5 = At least once a week \\
\bottomrule
\end{longtable}
\par\noindent{\footnotesize \textit{Note.} Response options and numeric codes reflect the human CCAM survey instrument \citep{leiserowitzClimateChangeAmerican2020}. The response options presented to the models were generated from the same codebook, so the nine items carrying a ``Don't know'' or ``Not sure'' category presented it to the models as well; the system prompt restricted responses to the range 1 through 6, which made the category unreachable on the six items coding it as 0 (see Methods and Appendix~\ref{app:matched}). The timing item is coded 1--6 in the human instrument and 0--5 in the prompt codebook, and model responses were shifted to the human coding before analysis. Question wording is abbreviated where the administered text is long: the policy items share a common stem (``How much do you support or oppose the following policies?''), and the item on whether global warming is happening was preceded by a definitional preamble.
}
}

\section{Descriptive Statistics}\label{app:descriptive}

\begin{table}[H]
\caption{\label{tab:demographic-distribution}Sample Demographic Distribution}
\begin{adjustbox}{center}
\begin{tabular}[t]{llr}
\toprule
Variable & Label (Value) & Count\\
\midrule
Sex/Gender & Male & 508\\
Sex/Gender & Female & 470\\
\midrule
Age & 18–34 & 204\\
Age & 35–54 & 309\\
Age & 55+ & 465\\
\midrule
Ideology & Very Liberal & 80\\
Ideology & Liberal & 199\\
Ideology & Moderate & 406\\
Ideology & Conservative & 204\\
Ideology & Very Conservative & 89\\
\midrule
Education & $<$ High School (1) & 321\\
Education & Some College (3) & 242\\
Education & Bachelor's + (4) & 415\\
\midrule
Income Category & $<$\$50k (1) & 200\\
Income Category & \$50k–\$99.9k (2) & 281\\
Income Category & \$100k+ (3) & 497\\
\midrule
Region & West & 231\\
Region & Northeast & 173\\
Region & Midwest & 206\\
Region & South & 368\\
\midrule
Religiosity & Not Religious (0) & 116\\
Religiosity & Religious (1) & 862\\
\midrule
Race/Ethnicity & White, non-Hispanic & 648\\
Race/Ethnicity & Black, non-Hispanic & 105\\
Race/Ethnicity & Other, non-Hispanic & 90\\
Race/Ethnicity & Hispanic & 135\\
\bottomrule
\end{tabular}
\end{adjustbox}
\end{table}

\begin{table}[H]

\centering
\caption{Respondent counts by race, political ideology, and gender (analytic sample, $N=978$).}\label{tab:cell_sizes}
\begin{tabular}{llrrr}
\toprule
Race & Ideology & Male & Female & Total \\
\midrule
White & Very liberal & 21 & 32 & 53 \\
White & Liberal & 73 & 65 & 138 \\
White & Moderate & 110 & 131 & 241 \\
White & Conservative & 92 & 60 & 152 \\
White & Very conservative & 41 & 23 & 64 \\
\addlinespace
Black & Very liberal & 8 & 5 & 13 \\
Black & Liberal & 8 & 13 & 21 \\
Black & Moderate & 30 & 23 & 53 \\
Black & Conservative & 7 & 6 & 13 \\
Black & Very conservative & 4 & 1 & 5 \\
\addlinespace
Other & Very liberal & 2 & 5 & 7 \\
Other & Liberal & 6 & 12 & 18 \\
Other & Moderate & 21 & 23 & 44 \\
Other & Conservative & 9 & 4 & 13 \\
Other & Very conservative & 5 & 3 & 8 \\
\addlinespace
Hispanic & Very liberal & 6 & 1 & 7 \\
Hispanic & Liberal & 10 & 12 & 22 \\
Hispanic & Moderate & 32 & 36 & 68 \\
Hispanic & Conservative & 16 & 10 & 26 \\
Hispanic & Very conservative & 7 & 5 & 12 \\
\midrule
\multicolumn{2}{l}{Total} & 508 & 470 & 978 \\
\bottomrule
\end{tabular}
\noindent\parbox{0.8\textwidth}{\vspace{0.5em}\small\textit{Note:} Cells at the very conservative extreme are small, particularly for Black ($n=5$; 1 female) and Hispanic ($n=12$) respondents, so the race-specific Female~$\times$~Very~Conservative interactions should be read as exploratory. Appendix~\ref{app:smallcell} reports four robustness checks on these estimates: a widened ideological band, correction for multiple comparisons, exact permutation tests, and partial pooling.}

\end{table}

\begin{table}[H]

\centering
\caption{Mean Scaled Climate Responses by Race and Gender: Humans versus LLM Predictions}\label{tab:gender_by_race}
\begin{tabular}[t]{lrrrrrrrr}
\toprule
\multicolumn{1}{c}{ } & \multicolumn{2}{c}{Respondents} & \multicolumn{3}{c}{Human} & \multicolumn{3}{c}{LLM} \\
\cmidrule(l{3pt}r{3pt}){2-3} \cmidrule(l{3pt}r{3pt}){4-6} \cmidrule(l{3pt}r{3pt}){7-9}
Race/Ethnicity & Male & Female & Male & Female & F $-$ M & Male & Female & F $-$ M\\
\midrule
White, non-Hispanic & 337 & 311 & 0.586 & 0.700 & 0.113 & 0.668 & 0.736 & 0.068\\
Black, non-Hispanic & 57 & 48 & 0.721 & 0.679 & -0.042 & 0.789 & 0.834 & 0.045\\
Other, non-Hispanic & 43 & 47 & 0.681 & 0.798 & 0.117 & 0.728 & 0.814 & 0.086\\
Hispanic & 71 & 64 & 0.649 & 0.733 & 0.084 & 0.719 & 0.754 & 0.034\\
\bottomrule
\end{tabular}
\noindent\parbox{0.8\textwidth}{\vspace{0.5em}\small\textit{Note:} Responses are min--max scaled to $[0,1]$ within each question, with higher values indicating more pro-climate responses, then averaged across the 20 items. Human means are computed once per respondent and question; LLM means pool the six models. F $-$ M is the female-minus-male difference. The human difference is positive for every racial group except Black respondents, for whom it reverses. The LLM difference is positive for all four groups.}

\end{table}

\begin{table}[H]

\centering
\caption{Mean LLM--human scaled gap by race, ideology, and gender (analytic sample, $N=978$)}\label{tab:gap_by_race_ideology}
\begin{tabular}{llrrrr}
\toprule
\multicolumn{3}{c}{ } & \multicolumn{3}{c}{Mean gap (LLM $-$ human)} \\
\cmidrule(l{3pt}r{3pt}){4-6}
Race/Ethnicity & Ideology & $n$ & Male & Female & Overall\\
\midrule
White, non-Hispanic & Very Liberal & 53 & 0.064 & 0.019 & 0.037\\
White, non-Hispanic & Liberal & 138 & 0.040 & 0.017 & 0.029\\
White, non-Hispanic & Moderate & 241 & 0.098 & 0.065 & 0.080\\
White, non-Hispanic & Conservative & 152 & 0.123 & 0.054 & 0.096\\
White, non-Hispanic & Very Conservative & 64 & 0.029 & -0.101 & -0.016\\
\addlinespace
Black, non-Hispanic & Very Liberal & 13 & 0.154 & 0.076 & 0.124\\
Black, non-Hispanic & Liberal & 21 & 0.090 & 0.167 & 0.137\\
Black, non-Hispanic & Moderate & 53 & 0.063 & 0.150 & 0.100\\
Black, non-Hispanic & Conservative & 13 & 0.176 & 0.231 & 0.201\\
Black, non-Hispanic & Very Conservative & 5 & -0.288 & 0.125 & -0.227\\
\addlinespace
Other, non-Hispanic & Very Liberal & 7 & 0.083 & 0.005 & 0.027\\
Other, non-Hispanic & Liberal & 18 & 0.048 & 0.024 & 0.032\\
Other, non-Hispanic & Moderate & 44 & 0.053 & 0.041 & 0.047\\
Other, non-Hispanic & Conservative & 13 & 0.117 & 0.037 & 0.092\\
Other, non-Hispanic & Very Conservative & 8 & -0.117 & -0.242 & -0.160\\
\addlinespace
Hispanic & Very Liberal & 7 & 0.038 & 0.018 & 0.036\\
Hispanic & Liberal & 22 & 0.046 & 0.007 & 0.025\\
Hispanic & Moderate & 68 & 0.087 & 0.061 & 0.074\\
Hispanic & Conservative & 26 & 0.033 & 0.008 & 0.023\\
Hispanic & Very Conservative & 12 & 0.142 & -0.202 & 0.001\\
\bottomrule
\end{tabular}
\noindent\parbox{0.8\textwidth}{\vspace{0.5em}\small\textit{Note:} Each cell is the mean scaled gap between LLM predictions and human responses ($\mathrm{LLM}-\mathrm{human}$), averaged across the 20 items; positive values indicate overestimation in the pro-climate direction. $n$ is the number of respondents; gender-split cells at the extremes rest on as few as one respondent and should be read as suggestive.}

\end{table}

\begin{table}[H]
\caption{\label{tab:desc_stats_human}Descriptive Statistics of Human Responses by Question}
\begin{adjustbox}{center}
\begin{tabular}[t]{lrrrrrr}
\toprule
\multicolumn{1}{c}{ } & \multicolumn{6}{c}{Descriptive Statistics} \\
\cmidrule(l{3pt}r{3pt}){2-7}
Question & $N$ & Mean & SD & Min & Max & Median\\
\midrule
cause\_recoded & 976 & 5.01 & 1.28 & 2.00 & 6.00 & 6.00\\
discuss\_GW & 977 & 2.12 & 0.93 & 1.00 & 4.00 & 2.00\\
fund\_research & 976 & 3.03 & 0.94 & 1.00 & 4.00 & 3.00\\
generate\_renewable & 977 & 2.96 & 0.93 & 1.00 & 4.00 & 3.00\\
happening & 977 & 2.59 & 0.73 & 1.00 & 3.00 & 3.00\\
\addlinespace
harm\_US & 904 & 2.87 & 1.04 & 1.00 & 4.00 & 3.00\\
harm\_dev\_countries & 872 & 3.13 & 1.06 & 1.00 & 4.00 & 4.00\\
harm\_future\_gen & 882 & 3.25 & 1.04 & 1.00 & 4.00 & 4.00\\
harm\_personally & 904 & 2.38 & 1.02 & 1.00 & 4.00 & 2.00\\
harm\_plants\_animals & 893 & 3.22 & 1.06 & 1.00 & 4.00 & 4.00\\
\addlinespace
hear\_GW\_media & 876 & 3.66 & 1.23 & 1.00 & 5.00 & 4.00\\
priority & 978 & 2.56 & 1.16 & 1.00 & 4.00 & 3.00\\
priority\_cleanenergy & 978 & 2.83 & 1.01 & 1.00 & 4.00 & 3.00\\
reduce\_tax & 974 & 2.78 & 1.01 & 1.00 & 4.00 & 3.00\\
reg\_CO2\_pollutant & 975 & 2.92 & 0.96 & 1.00 & 4.00 & 3.00\\
\addlinespace
sci\_consensus & 977 & 3.06 & 1.21 & 1.00 & 4.00 & 4.00\\
teach\_gw & 978 & 3.07 & 1.02 & 1.00 & 4.00 & 3.00\\
transition\_economy & 974 & 2.71 & 1.08 & 1.00 & 4.00 & 3.00\\
when\_harm\_US & 975 & 4.31 & 1.93 & 1.00 & 6.00 & 5.00\\
worry & 978 & 2.75 & 1.03 & 1.00 & 4.00 & 3.00\\
\bottomrule
\end{tabular}
\end{adjustbox}
\end{table}

\begin{longtable}{llrrrrrr}
\caption{\label{tab:desc_stats_question_model}Descriptive Statistics by Question and LLM Model}\\
\toprule
\multicolumn{2}{c}{ } & \multicolumn{6}{c}{Descriptive Statistics} \\
\cmidrule(l{3pt}r{3pt}){3-8}
Question & Model & $N$ & Mean & SD & Min & Max & Median\\
\midrule
\endfirsthead

\caption[]{Descriptive Statistics by Question and LLM Model (continued)}\\
\toprule
\multicolumn{2}{c}{ } & \multicolumn{6}{c}{Descriptive Statistics} \\
\cmidrule(l{3pt}r{3pt}){3-8}
Question & Model & $N$ & Mean & SD & Min & Max & Median\\
\midrule
\endhead

\midrule
\multicolumn{8}{r}{(continued on next page)}\\
\endfoot

\bottomrule
\endlastfoot

\textbf{cause\_recoded} & \textbf{Total} & \textbf{5,856} & \textbf{5.22} & \textbf{1.15} & \textbf{1.00} & \textbf{6.00} & \textbf{6.00}\\
cause\_recoded & gemma3 & 976 & 4.70 & 1.40 & 2.00 & 6.00 & 5.50\\
cause\_recoded & gpt-4o-2024-08-06 & 976 & 5.17 & 0.84 & 1.00 & 6.00 & 5.00\\
cause\_recoded & gpt-5-2025-08-07 & 976 & 4.37 & 1.55 & 1.00 & 6.00 & 5.00\\
cause\_recoded & llama3 & 976 & 5.99 & 0.11 & 4.00 & 6.00 & 6.00\\
\addlinespace
cause\_recoded & mistral & 976 & 5.90 & 0.55 & 1.00 & 6.00 & 6.00\\
cause\_recoded & phi4 & 976 & 5.18 & 0.75 & 1.00 & 6.00 & 5.00\\
\textbf{discuss\_GW} & \textbf{Total} & \textbf{5,837} & \textbf{2.67} & \textbf{0.80} & \textbf{1.00} & \textbf{4.00} & \textbf{3.00}\\
discuss\_GW & gemma3 & 964 & 2.12 & 0.60 & 1.00 & 3.00 & 2.00\\
discuss\_GW & gpt-4o-2024-08-06 & 977 & 2.55 & 0.66 & 1.00 & 4.00 & 2.00\\
\addlinespace
discuss\_GW & gpt-5-2025-08-07 & 966 & 2.83 & 0.76 & 1.00 & 4.00 & 3.00\\
discuss\_GW & llama3 & 977 & 2.86 & 0.35 & 2.00 & 3.00 & 3.00\\
discuss\_GW & mistral & 976 & 2.55 & 1.16 & 1.00 & 4.00 & 3.00\\
discuss\_GW & phi4 & 977 & 3.13 & 0.62 & 1.00 & 4.00 & 3.00\\
\textbf{fund\_research} & \textbf{Total} & \textbf{5,810} & \textbf{3.28} & \textbf{0.73} & \textbf{1.00} & \textbf{4.00} & \textbf{3.00}\\
\addlinespace
fund\_research & gemma3 & 963 & 2.63 & 0.74 & 1.00 & 4.00 & 3.00\\
fund\_research & gpt-4o-2024-08-06 & 976 & 3.52 & 0.59 & 1.00 & 4.00 & 4.00\\
fund\_research & gpt-5-2025-08-07 & 965 & 3.31 & 0.82 & 1.00 & 4.00 & 4.00\\
fund\_research & llama3 & 955 & 3.01 & 0.19 & 2.00 & 4.00 & 3.00\\
fund\_research & mistral & 975 & 3.75 & 0.62 & 1.00 & 4.00 & 4.00\\
\addlinespace
fund\_research & phi4 & 976 & 3.45 & 0.66 & 1.00 & 4.00 & 4.00\\
\textbf{generate\_renewable} & \textbf{Total} & \textbf{5,810} & \textbf{3.12} & \textbf{0.83} & \textbf{1.00} & \textbf{4.00} & \textbf{3.00}\\
generate\_renewable & gemma3 & 964 & 2.41 & 0.77 & 1.00 & 4.00 & 3.00\\
generate\_renewable & gpt-4o-2024-08-06 & 977 & 3.53 & 0.58 & 1.00 & 4.00 & 4.00\\
generate\_renewable & gpt-5-2025-08-07 & 966 & 3.29 & 0.83 & 1.00 & 4.00 & 3.50\\
\addlinespace
generate\_renewable & llama3 & 950 & 2.90 & 0.30 & 2.00 & 4.00 & 3.00\\
generate\_renewable & mistral & 976 & 3.12 & 1.07 & 1.00 & 4.00 & 4.00\\
generate\_renewable & phi4 & 977 & 3.45 & 0.68 & 1.00 & 4.00 & 4.00\\
\textbf{happening} & \textbf{Total} & \textbf{5,723} & \textbf{2.66} & \textbf{0.69} & \textbf{1.00} & \textbf{3.00} & \textbf{3.00}\\
happening & gemma3 & 964 & 1.98 & 0.81 & 1.00 & 3.00 & 2.00\\
\addlinespace
happening & gpt-4o-2024-08-06 & 977 & 2.72 & 0.67 & 1.00 & 3.00 & 3.00\\
happening & gpt-5-2025-08-07 & 854 & 2.71 & 0.60 & 1.00 & 3.00 & 3.00\\
happening & llama3 & 977 & 3.00 & 0.00 & 3.00 & 3.00 & 3.00\\
happening & mistral & 974 & 2.69 & 0.73 & 1.00 & 3.00 & 3.00\\
happening & phi4 & 977 & 2.87 & 0.50 & 1.00 & 3.00 & 3.00\\
\addlinespace
\textbf{harm\_US} & \textbf{Total} & \textbf{5,402} & \textbf{3.10} & \textbf{0.78} & \textbf{1.00} & \textbf{4.00} & \textbf{3.00}\\
harm\_US & gemma3 & 893 & 2.59 & 0.64 & 1.00 & 4.00 & 3.00\\
harm\_US & gpt-4o-2024-08-06 & 904 & 3.43 & 0.68 & 2.00 & 4.00 & 4.00\\
harm\_US & gpt-5-2025-08-07 & 894 & 3.19 & 0.87 & 1.00 & 4.00 & 3.00\\
harm\_US & llama3 & 904 & 2.67 & 0.50 & 1.00 & 4.00 & 3.00\\
\addlinespace
harm\_US & mistral & 903 & 3.46 & 0.81 & 1.00 & 4.00 & 4.00\\
harm\_US & phi4 & 904 & 3.23 & 0.65 & 1.00 & 4.00 & 3.00\\
\textbf{harm\_dev\_countries} & \textbf{Total} & \textbf{5,211} & \textbf{3.32} & \textbf{0.76} & \textbf{1.00} & \textbf{4.00} & \textbf{3.00}\\
harm\_dev\_countries & gemma3 & 862 & 2.38 & 0.59 & 1.00 & 4.00 & 2.00\\
harm\_dev\_countries & gpt-4o-2024-08-06 & 872 & 3.66 & 0.59 & 2.00 & 4.00 & 4.00\\
\addlinespace
harm\_dev\_countries & gpt-5-2025-08-07 & 862 & 3.30 & 0.85 & 1.00 & 4.00 & 4.00\\
harm\_dev\_countries & llama3 & 872 & 2.93 & 0.26 & 2.00 & 3.00 & 3.00\\
harm\_dev\_countries & mistral & 871 & 3.98 & 0.18 & 1.00 & 4.00 & 4.00\\
harm\_dev\_countries & phi4 & 872 & 3.67 & 0.48 & 1.00 & 4.00 & 4.00\\
\textbf{harm\_future\_gen} & \textbf{Total} & \textbf{5,270} & \textbf{3.32} & \textbf{0.78} & \textbf{1.00} & \textbf{4.00} & \textbf{3.00}\\
\addlinespace
harm\_future\_gen & gemma3 & 871 & 2.57 & 0.76 & 1.00 & 4.00 & 3.00\\
harm\_future\_gen & gpt-4o-2024-08-06 & 882 & 3.64 & 0.62 & 1.00 & 4.00 & 4.00\\
harm\_future\_gen & gpt-5-2025-08-07 & 872 & 3.32 & 0.84 & 1.00 & 4.00 & 4.00\\
harm\_future\_gen & llama3 & 882 & 2.87 & 0.34 & 2.00 & 3.00 & 3.00\\
harm\_future\_gen & mistral & 881 & 3.95 & 0.36 & 1.00 & 4.00 & 4.00\\
\addlinespace
harm\_future\_gen & phi4 & 882 & 3.55 & 0.69 & 1.00 & 4.00 & 4.00\\
\textbf{harm\_personally} & \textbf{Total} & \textbf{5,402} & \textbf{2.62} & \textbf{0.79} & \textbf{1.00} & \textbf{4.00} & \textbf{3.00}\\
harm\_personally & gemma3 & 893 & 2.17 & 0.75 & 1.00 & 4.00 & 2.00\\
harm\_personally & gpt-4o-2024-08-06 & 904 & 2.58 & 0.56 & 1.00 & 4.00 & 3.00\\
harm\_personally & gpt-5-2025-08-07 & 894 & 2.82 & 0.77 & 1.00 & 4.00 & 3.00\\
\addlinespace
harm\_personally & llama3 & 904 & 2.53 & 0.55 & 1.00 & 3.00 & 3.00\\
harm\_personally & mistral & 903 & 2.82 & 1.09 & 1.00 & 4.00 & 3.00\\
harm\_personally & phi4 & 904 & 2.82 & 0.69 & 1.00 & 4.00 & 3.00\\
\textbf{harm\_plants\_animals} & \textbf{Total} & \textbf{5,336} & \textbf{3.30} & \textbf{0.79} & \textbf{1.00} & \textbf{4.00} & \textbf{3.00}\\
harm\_plants\_animals & gemma3 & 882 & 2.59 & 0.66 & 1.00 & 4.00 & 3.00\\
\addlinespace
harm\_plants\_animals & gpt-4o-2024-08-06 & 893 & 3.67 & 0.59 & 1.00 & 4.00 & 4.00\\
harm\_plants\_animals & gpt-5-2025-08-07 & 883 & 3.30 & 0.85 & 1.00 & 4.00 & 4.00\\
harm\_plants\_animals & llama3 & 893 & 2.77 & 0.42 & 2.00 & 3.00 & 3.00\\
harm\_plants\_animals & mistral & 892 & 3.81 & 0.62 & 1.00 & 4.00 & 4.00\\
harm\_plants\_animals & phi4 & 893 & 3.64 & 0.62 & 1.00 & 4.00 & 4.00\\
\addlinespace
\textbf{hear\_GW\_media} & \textbf{Total} & \textbf{5,255} & \textbf{3.64} & \textbf{1.17} & \textbf{1.00} & \textbf{5.00} & \textbf{4.00}\\
hear\_GW\_media & gemma3 & 876 & 2.21 & 0.64 & 1.00 & 5.00 & 2.00\\
hear\_GW\_media & gpt-4o-2024-08-06 & 876 & 5.00 & 0.00 & 5.00 & 5.00 & 5.00\\
hear\_GW\_media & gpt-5-2025-08-07 & 875 & 4.04 & 1.21 & 1.00 & 5.00 & 5.00\\
hear\_GW\_media & llama3 & 876 & 3.06 & 0.41 & 2.00 & 4.00 & 3.00\\
\addlinespace
hear\_GW\_media & mistral & 876 & 3.98 & 1.16 & 1.00 & 5.00 & 4.00\\
hear\_GW\_media & phi4 & 876 & 3.54 & 0.52 & 1.00 & 5.00 & 4.00\\
\textbf{priority} & \textbf{Total} & \textbf{5,842} & \textbf{2.93} & \textbf{0.97} & \textbf{1.00} & \textbf{4.00} & \textbf{3.00}\\
priority & gemma3 & 965 & 2.19 & 0.56 & 1.00 & 4.00 & 2.00\\
priority & gpt-4o-2024-08-06 & 978 & 2.73 & 1.20 & 1.00 & 4.00 & 3.00\\
\addlinespace
priority & gpt-5-2025-08-07 & 967 & 2.95 & 0.95 & 1.00 & 4.00 & 3.00\\
priority & llama3 & 977 & 3.11 & 0.31 & 3.00 & 4.00 & 3.00\\
priority & mistral & 977 & 3.28 & 1.09 & 1.00 & 4.00 & 4.00\\
priority & phi4 & 978 & 3.30 & 0.93 & 1.00 & 4.00 & 4.00\\
\textbf{priority\_cleanenergy} & \textbf{Total} & \textbf{5,843} & \textbf{3.34} & \textbf{0.71} & \textbf{1.00} & \textbf{4.00} & \textbf{3.00}\\
\addlinespace
priority\_cleanenergy & gemma3 & 965 & 2.82 & 0.52 & 1.00 & 4.00 & 3.00\\
priority\_cleanenergy & gpt-4o-2024-08-06 & 978 & 3.37 & 0.78 & 1.00 & 4.00 & 4.00\\
priority\_cleanenergy & gpt-5-2025-08-07 & 967 & 3.22 & 0.84 & 1.00 & 4.00 & 3.00\\
priority\_cleanenergy & llama3 & 978 & 3.11 & 0.31 & 3.00 & 4.00 & 3.00\\
priority\_cleanenergy & mistral & 977 & 3.89 & 0.33 & 1.00 & 4.00 & 4.00\\
\addlinespace
priority\_cleanenergy & phi4 & 978 & 3.59 & 0.71 & 1.00 & 4.00 & 4.00\\
\textbf{reduce\_tax} & \textbf{Total} & \textbf{5,795} & \textbf{2.67} & \textbf{0.97} & \textbf{1.00} & \textbf{4.00} & \textbf{3.00}\\
reduce\_tax & gemma3 & 961 & 1.97 & 0.68 & 1.00 & 3.00 & 2.00\\
reduce\_tax & gpt-4o-2024-08-06 & 974 & 2.82 & 0.90 & 1.00 & 4.00 & 3.00\\
reduce\_tax & gpt-5-2025-08-07 & 963 & 2.98 & 0.86 & 1.00 & 4.00 & 3.00\\
\addlinespace
reduce\_tax & llama3 & 950 & 2.61 & 0.50 & 2.00 & 4.00 & 3.00\\
reduce\_tax & mistral & 973 & 2.68 & 1.33 & 1.00 & 4.00 & 3.00\\
reduce\_tax & phi4 & 974 & 2.92 & 0.96 & 1.00 & 4.00 & 3.00\\
\textbf{reg\_CO2\_pollutant} & \textbf{Total} & \textbf{5,818} & \textbf{2.81} & \textbf{1.01} & \textbf{1.00} & \textbf{4.00} & \textbf{3.00}\\
reg\_CO2\_pollutant & gemma3 & 962 & 1.98 & 0.81 & 1.00 & 3.00 & 2.00\\
\addlinespace
reg\_CO2\_pollutant & gpt-4o-2024-08-06 & 975 & 3.03 & 1.02 & 1.00 & 4.00 & 3.00\\
reg\_CO2\_pollutant & gpt-5-2025-08-07 & 964 & 3.09 & 0.94 & 1.00 & 4.00 & 3.00\\
reg\_CO2\_pollutant & llama3 & 968 & 2.68 & 0.56 & 2.00 & 4.00 & 3.00\\
reg\_CO2\_pollutant & mistral & 974 & 3.02 & 1.26 & 1.00 & 4.00 & 4.00\\
reg\_CO2\_pollutant & phi4 & 975 & 3.03 & 0.87 & 1.00 & 4.00 & 3.00\\
\addlinespace
\textbf{sci\_consensus} & \textbf{Total} & \textbf{5,837} & \textbf{3.51} & \textbf{0.84} & \textbf{1.00} & \textbf{4.00} & \textbf{4.00}\\
sci\_consensus & gemma3 & 964 & 2.82 & 0.92 & 2.00 & 4.00 & 2.00\\
sci\_consensus & gpt-4o-2024-08-06 & 977 & 3.55 & 0.84 & 1.00 & 4.00 & 4.00\\
sci\_consensus & gpt-5-2025-08-07 & 966 & 3.22 & 0.94 & 1.00 & 4.00 & 4.00\\
sci\_consensus & llama3 & 977 & 3.95 & 0.32 & 2.00 & 4.00 & 4.00\\
\addlinespace
sci\_consensus & mistral & 976 & 3.65 & 0.75 & 1.00 & 4.00 & 4.00\\
sci\_consensus & phi4 & 977 & 3.88 & 0.48 & 1.00 & 4.00 & 4.00\\
\textbf{teach\_gw} & \textbf{Total} & \textbf{5,832} & \textbf{3.26} & \textbf{0.81} & \textbf{1.00} & \textbf{4.00} & \textbf{3.00}\\
teach\_gw & gemma3 & 965 & 2.51 & 0.62 & 1.00 & 4.00 & 2.00\\
teach\_gw & gpt-4o-2024-08-06 & 978 & 3.26 & 0.93 & 1.00 & 4.00 & 4.00\\
\addlinespace
teach\_gw & gpt-5-2025-08-07 & 967 & 3.24 & 0.85 & 1.00 & 4.00 & 3.00\\
teach\_gw & llama3 & 967 & 3.18 & 0.51 & 2.00 & 4.00 & 3.00\\
teach\_gw & mistral & 977 & 3.84 & 0.56 & 1.00 & 4.00 & 4.00\\
teach\_gw & phi4 & 978 & 3.50 & 0.67 & 1.00 & 4.00 & 4.00\\
\textbf{transition\_economy} & \textbf{Total} & \textbf{5,809} & \textbf{2.88} & \textbf{0.99} & \textbf{1.00} & \textbf{4.00} & \textbf{3.00}\\
\addlinespace
transition\_economy & gemma3 & 961 & 2.15 & 0.85 & 1.00 & 4.00 & 2.00\\
transition\_economy & gpt-4o-2024-08-06 & 974 & 3.12 & 0.97 & 1.00 & 4.00 & 3.00\\
transition\_economy & gpt-5-2025-08-07 & 963 & 3.11 & 0.93 & 1.00 & 4.00 & 3.00\\
transition\_economy & llama3 & 964 & 2.89 & 0.66 & 2.00 & 4.00 & 3.00\\
transition\_economy & mistral & 973 & 2.90 & 1.24 & 1.00 & 4.00 & 4.00\\
\addlinespace
transition\_economy & phi4 & 974 & 3.10 & 0.85 & 1.00 & 4.00 & 3.00\\
\textbf{when\_harm\_US} & \textbf{Total} & \textbf{5,849} & \textbf{4.42} & \textbf{1.56} & \textbf{1.00} & \textbf{6.00} & \textbf{5.00}\\
when\_harm\_US & gemma3 & 975 & 3.44 & 2.03 & 1.00 & 6.00 & 3.00\\
when\_harm\_US & gpt-4o-2024-08-06 & 975 & 4.95 & 1.52 & 1.00 & 6.00 & 6.00\\
when\_harm\_US & gpt-5-2025-08-07 & 974 & 4.64 & 1.34 & 2.00 & 6.00 & 5.00\\
\addlinespace
when\_harm\_US & llama3 & 975 & 3.58 & 0.93 & 1.00 & 5.00 & 3.00\\
when\_harm\_US & mistral & 975 & 4.59 & 1.06 & 1.00 & 6.00 & 5.00\\
when\_harm\_US & phi4 & 975 & 5.30 & 1.23 & 1.00 & 6.00 & 6.00\\
\textbf{worry} & \textbf{Total} & \textbf{5,836} & \textbf{3.01} & \textbf{0.88} & \textbf{1.00} & \textbf{4.00} & \textbf{3.00}\\
worry & gemma3 & 965 & 2.74 & 0.72 & 1.00 & 4.00 & 3.00\\
\addlinespace
worry & gpt-4o-2024-08-06 & 978 & 2.98 & 0.96 & 1.00 & 4.00 & 3.00\\
worry & gpt-5-2025-08-07 & 967 & 3.07 & 0.90 & 1.00 & 4.00 & 3.00\\
worry & llama3 & 971 & 2.83 & 0.50 & 2.00 & 4.00 & 3.00\\
worry & mistral & 977 & 3.13 & 1.18 & 1.00 & 4.00 & 4.00\\
worry & phi4 & 978 & 3.31 & 0.72 & 1.00 & 4.00 & 3.00\\
\end{longtable}

\begin{table}[H]
\caption{\label{tab:excluded_vs_retained}Demographic Comparison: Excluded vs. Retained Respondents}
\begin{adjustbox}{center}
\begin{tabular}[t]{llrr}
\toprule
Variable & Category & Excluded (N=22) & Retained (N=978)\\
\midrule
Sex/Gender & Male & 7 (31.8\%) & 508 (51.9\%)\\
Sex/Gender & Female & 15 (68.2\%) & 470 (48.1\%)\\
Race/Ethnicity & White, non-Hispanic & 11 (50\%) & 648 (66.3\%)\\
Race/Ethnicity & Black, non-Hispanic & 8 (36.4\%) & 105 (10.7\%)\\
Race/Ethnicity & Other, non-Hispanic & 1 (4.5\%) & 90 (9.2\%)\\
\addlinespace
Race/Ethnicity & Hispanic & 2 (9.1\%) & 135 (13.8\%)\\
Political Ideology & Very Liberal & 0 (0\%) & 80 (8.2\%)\\
Political Ideology & Liberal & 0 (0\%) & 199 (20.3\%)\\
Political Ideology & Moderate & 0 (0\%) & 406 (41.5\%)\\
Political Ideology & Conservative & 0 (0\%) & 204 (20.9\%)\\
\addlinespace
Political Ideology & Very Conservative & 0 (0\%) & 89 (9.1\%)\\
Political Ideology & Missing & 22 (100\%) & 0 (0\%)\\
Education & < High School & 12 (54.5\%) & 321 (32.8\%)\\
Education & High School & 0 (0\%) & 0 (0\%)\\
Education & Some College & 4 (18.2\%) & 242 (24.7\%)\\
\addlinespace
Education & Bachelor's+ & 6 (27.3\%) & 415 (42.4\%)\\
Income & <\$50k & 7 (31.8\%) & 200 (20.4\%)\\
Income & \$50k-\$99.9k & 8 (36.4\%) & 281 (28.7\%)\\
Income & \$100k+ & 7 (31.8\%) & 497 (50.8\%)\\
Age Category & 18-34 & 5 (22.7\%) & 204 (20.9\%)\\
\addlinespace
Age Category & 35-54 & 6 (27.3\%) & 309 (31.6\%)\\
Age Category & 55+ & 11 (50\%) & 465 (47.5\%)\\
Region & Northeast & 4 (18.2\%) & 173 (17.7\%)\\
Region & Midwest & 1 (4.5\%) & 206 (21.1\%)\\
Region & South & 13 (59.1\%) & 368 (37.6\%)\\
\addlinespace
Region & West & 4 (18.2\%) & 231 (23.6\%)\\
Religiosity & Religious & 21 (95.5\%) & 862 (88.1\%)\\
Religiosity & Not Religious & 1 (4.5\%) & 116 (11.9\%)\\
\bottomrule
\end{tabular}
\end{adjustbox}
\noindent\parbox{0.8\textwidth}{\vspace{0.5em}\small\textit{Note:} Respondents were excluded only if political ideology was missing (``don't know''). Item-level blank and refused entries, together with ``don't know'' responses on the items where the CCAM instrument codes that category outside the substantive response scale, were treated as missing and dropped at the question level rather than by excluding the respondent. Percentages are within-group column percentages.}
\end{table}

\subsection{Variance Analysis}\label{app:variance}

To quantify the degree to which LLMs compress the range of predicted climate opinions, we calculated variance ratios (LLM variance / Human variance) for each of the 20 survey questions (Table~\ref{tab:variance}). A variance ratio below 1 indicates that LLM predictions exhibit less variability than actual human responses. Across all 20 questions, 19 showed variance ratios below 1, with a mean ratio of 0.71. This indicates that LLM predictions exhibited 29\% less variance than human responses on average. Compression was most pronounced for questions about scientific consensus (ratio = 0.48), clean energy prioritization (ratio = 0.49), and harm to developing countries (ratio = 0.52). Only one question---regulating CO2 as a pollutant---showed a ratio slightly above 1 (ratio = 1.10). \begin{table}[H]
\caption{\label{tab:variance}Variance Compression by Question.}
\begin{adjustbox}{width=0.8\textwidth, center}
\begin{tabular}{lrrrr}
\toprule
Question & Human Variance & LLM Variance & Variance Ratio & Compression (\%)\\
\midrule
cause\_recoded & 0.0659 & 0.0532 & 0.807 & 19.3\\
discuss\_GW & 0.0954 & 0.0707 & 0.741 & 25.9\\
fund\_research & 0.0987 & 0.0596 & 0.604 & 39.6\\
generate\_renewable & 0.0966 & 0.0774 & 0.801 & 19.9\\
happening & 0.1337 & 0.1196 & 0.894 & 10.6\\
\addlinespace
harm\_dev\_countries & 0.0700 & 0.0361 & 0.515 & 48.5\\
harm\_future\_gen & 0.0679 & 0.0383 & 0.565 & 43.5\\
harm\_personally & 0.0647 & 0.0393 & 0.607 & 39.3\\
harm\_plants\_animals & 0.0698 & 0.0390 & 0.558 & 44.2\\
harm\_US & 0.0676 & 0.0381 & 0.563 & 43.7\\
\addlinespace
hear\_GW\_media & 0.0607 & 0.0545 & 0.899 & 10.1\\
priority & 0.1501 & 0.1053 & 0.702 & 29.8\\
priority\_cleanenergy & 0.1132 & 0.0555 & 0.490 & 51.0\\
reduce\_tax & 0.1131 & 0.1042 & 0.921 & 7.9\\
reg\_CO2\_pollutant & 0.1029 & 0.1137 & 1.105 & -10.5\\
\addlinespace
sci\_consensus & 0.1613 & 0.0780 & 0.484 & 51.6\\
teach\_gw & 0.1148 & 0.0735 & 0.640 & 36.0\\
transition\_economy & 0.1289 & 0.1091 & 0.846 & 15.4\\
when\_harm\_US & 0.1493 & 0.0968 & 0.648 & 35.2\\
worry & 0.1182 & 0.0854 & 0.722 & 27.8\\
\addlinespace
\textbf{Mean} & \textbf{0.1021} & \textbf{0.0724} & \textbf{0.706} & \textbf{29.4}\\
\bottomrule
\end{tabular}   
\end{adjustbox}
\noindent\parbox{0.8\textwidth}{\vspace{0.5em}\small\textit{Note:} Variance ratio $=$ LLM variance / Human variance. Values below 1 indicate LLM predictions are more compressed than human responses. Compression percentage $=$ $(1 - \text{variance ratio}) \times 100$.}
\end{table}

\section{Statistical Models}
\begin{table}[H]
\caption{Fixed Effects Models with Question Controls}\label{tab:tab_primary}
\begin{adjustbox}{center}
\begin{tabular}{lccc}       \tabularnewline \midrule \midrule                                           & LLM - Human    & Human          & LLM \\          Model:                              & (1)            & (2)            & (3)\\         \midrule       \emph{Variables}\\       Female                              & -0.030$^{***}$ & 0.062$^{***}$  & 0.032$^{***}$\\                                              & (0.011)        & (0.012)        & (0.005)\\          Age                                 & -0.0006$^{*}$  & 0.0002         & -0.0003$^{**}$\\                                              & (0.0003)       & (0.0003)       & (0.0001)\\          Race/Ethnicity: Black               & 0.046$^{**}$   & 0.038$^{*}$    & 0.084$^{***}$\\                                              & (0.018)        & (0.019)        & (0.009)\\          Race/Ethnicity: Other, non-Hispanic & -0.033$^{*}$   & 0.081$^{***}$  & 0.048$^{***}$\\                                              & (0.017)        & (0.019)        & (0.008)\\          Race/Ethnicity: Hispanic            & -0.015         & 0.065$^{***}$  & 0.050$^{***}$\\                                              & (0.017)        & (0.020)        & (0.008)\\          Ideology: Very Liberal              & -0.031$^{*}$   & 0.169$^{***}$  & 0.138$^{***}$\\                                              & (0.017)        & (0.018)        & (0.007)\\          Ideology: Liberal                   & -0.041$^{***}$ & 0.134$^{***}$  & 0.094$^{***}$\\                                              & (0.013)        & (0.015)        & (0.006)\\          Ideology: Conservative              & 0.006          & -0.182$^{***}$ & -0.176$^{***}$\\                                              & (0.016)        & (0.018)        & (0.007)\\          Ideology: Very Conservative         & -0.124$^{***}$ & -0.268$^{***}$ & -0.392$^{***}$\\                                              & (0.023)        & (0.024)        & (0.008)\\          Education                           & 0.007          & 0.016$^{***}$  & 0.023$^{***}$\\                                              & (0.005)        & (0.005)        & (0.002)\\          Income                              & 0.025$^{***}$  & -0.012         & 0.013$^{***}$\\                                              & (0.008)        & (0.008)        & (0.004)\\          Region: Northeast                   & -0.006         & 0.008          & 0.002\\                                              & (0.016)        & (0.018)        & (0.007)\\          Region: Midwest                     & -0.008         & 0.008          & 0.0006\\                                              & (0.016)        & (0.017)        & (0.007)\\          Region: South                       & -0.003         & -0.003         & -0.006\\                                              & (0.014)        & (0.015)        & (0.006)\\          Religious                           & 0.027$^{*}$    & -0.059$^{***}$ & -0.032$^{***}$\\                                              & (0.015)        & (0.016)        & (0.006)\\          \midrule       \emph{Fixed-effects}\\       Questions                           & Yes            & Yes            & Yes\\         LLMs                                & Yes            & Yes            & Yes\\         \midrule       \emph{Fit statistics}\\       Observations                        & 113,373        & 113,373        & 113,373\\         R$^2$                               & 0.14068        & 0.31665        & 0.55547\\         Within R$^2$                        & 0.02570        & 0.23908        & 0.42347\\         \midrule \midrule       \multicolumn{4}{l}{\emph{Clustered (case\_id) standard-errors in parentheses}}\\       \multicolumn{4}{l}{\emph{Signif. Codes: ***: 0.01, **: 0.05, *: 0.1}}\\    \end{tabular}
\end{adjustbox}
\noindent\parbox{0.8\textwidth}{\vspace{0.5em}\small\textit{Note:} The dependent variable is the normalized gap between LLM and human responses (LLM—Human), where positive values indicate LLM overestimation. Reference categories: Male, White non-Hispanic, Moderate (ideology), West (region). Clustered (\texttt{case\_id}) standard errors in parentheses. Fixed effects for question and LLM model included. Significance: $*** p<.01, ** p<.05, * p<.1$.}
\end{table}

\begin{table}[H]
\caption{Gender $\times$ Ideology Interaction Models by Racial Group}\label{tab:tab_racial_subset}
\begin{adjustbox}{width=0.8\textwidth, center}
\begin{tabular}{lcccc}       \tabularnewline \midrule \midrule                                                    & White          & Black          & Hispanic       & Other \\          Model:                                       & (1)            & (2)            & (3)            & (4)\\         \midrule       \emph{Variables}\\       Female                                       & -0.026         & 0.069          & -0.008         & -0.011\\                                                       & (0.023)        & (0.043)        & (0.041)        & (0.044)\\          Ideology: Very Liberal                       & -0.040         & 0.094          & -0.021         & 0.089\\                                                       & (0.036)        & (0.066)        & (0.060)        & (0.059)\\          Ideology: Liberal                            & -0.059$^{**}$  & 0.025          & -0.025         & -0.0003\\                                                       & (0.026)        & (0.036)        & (0.046)        & (0.066)\\          Ideology: Conservative                       & 0.026          & 0.090          & -0.065         & 0.068\\                                                       & (0.026)        & (0.059)        & (0.059)        & (0.068)\\          Ideology: Very Conservative                  & -0.068$^{**}$  & -0.362$^{***}$ & 0.037          & -0.197$^{**}$\\                                                       & (0.028)        & (0.091)        & (0.043)        & (0.087)\\          Age                                          & -0.0008$^{**}$ & 0.0003         & -0.0007        & $-9.26\times 10^{-5}$\\                                                        & (0.0004)       & (0.0008)       & (0.001)        & (0.0009)\\          Education                                    & 0.006          & 0.005          & 0.003          & 0.003\\                                                       & (0.006)        & (0.014)        & (0.014)        & (0.014)\\          Region: Northeast                            & -0.011         & 0.065          & 0.008          & -0.040\\                                                       & (0.020)        & (0.045)        & (0.040)        & (0.043)\\          Region: Midwest                              & -0.018         & 0.070          & -0.009         & 0.057\\                                                       & (0.019)        & (0.054)        & (0.062)        & (0.047)\\          Region: South                                & -0.014         & 0.065          & 0.001          & 0.0008\\                                                       & (0.018)        & (0.043)        & (0.034)        & (0.039)\\          Income                                       & 0.027$^{***}$  & 0.001          & 0.045$^{**}$   & 0.027\\                                                       & (0.009)        & (0.021)        & (0.023)        & (0.026)\\          Religious                                    & 0.007          & 0.056          & 0.079$^{**}$   & 0.082$^{**}$\\                                                       & (0.019)        & (0.045)        & (0.035)        & (0.034)\\          Female $\times$ Ideology: Very Liberal       & -0.026         & -0.133         & -0.012         & -0.087\\                                                       & (0.040)        & (0.111)        & (0.070)        & (0.070)\\          Female $\times$ Ideology: Liberal            & 0.006          & 0.015          & -0.041         & -0.011\\                                                       & (0.032)        & (0.062)        & (0.056)        & (0.077)\\          Female $\times$ Ideology: Conservative       & -0.045         & -0.014         & -0.009         & -0.102\\                                                       & (0.040)        & (0.103)        & (0.100)        & (0.080)\\          Female $\times$ Ideology: Very Conservative  & -0.096$^{*}$   & 0.323$^{***}$  & -0.319$^{***}$ & -0.097\\                                                       & (0.054)        & (0.111)        & (0.116)        & (0.141)\\          \midrule       \emph{Fixed-effects}\\       Questions                                    & Yes            & Yes            & Yes            & Yes\\         LLMs                                         & Yes            & Yes            & Yes            & Yes\\         \midrule       \emph{Fit statistics}\\       Observations                                 & 75,224         & 11,888         & 15,742         & 10,519\\         R$^2$                                        & 0.14273        & 0.22931        & 0.18257        & 0.18384\\         Within R$^2$                                 & 0.02932        & 0.09756        & 0.04756        & 0.06863\\         \midrule \midrule       \multicolumn{5}{l}{\emph{Clustered (case\_id) standard-errors in parentheses}}\\       \multicolumn{5}{l}{\emph{Signif. Codes: ***: 0.01, **: 0.05, *: 0.1}}\\    \end{tabular}
\end{adjustbox}
\noindent\parbox{0.8\textwidth}{\vspace{0.5em}\small\textit{Note:} Separate models estimated for each racial group. The dependent variable is the normalized gap between LLM and human responses (LLM -- Human), where positive values indicate LLM overestimation. Models include Gender $\times$ Ideology interaction to test whether this relationship—central to the three-way interaction examined in the main analysis—varies within racial groups. Reference categories: Male, Moderate (ideology), West (region). Clustered (\texttt{case\_id}) standard errors in parentheses. Fixed effects for question and LLM model included. Significance: $*** p < .01, ** p < .05, * p < .1$.}
\end{table}

\begin{table}[H]
\caption{Gender $\times$ Ideology Interaction Models by Racial Group, Conservative and Very Conservative Collapsed}\label{tab:tab_racial_subset_collapsed}
\begin{adjustbox}{width=0.8\textwidth, center}
\begin{tabular}{lcccc}       \tabularnewline \midrule \midrule                                                         & White          & Black       & Hispanic     & Other \\          Model:                                            & (1)            & (2)         & (3)          & (4)\\         \midrule       \emph{Variables}\\       Female                                            & -0.025         & 0.072       & -0.008       & -0.009\\                                                            & (0.023)        & (0.044)     & (0.041)      & (0.045)\\          demo\_ideology\_cVeryLiberal                      & -0.041         & 0.107       & -0.023       & 0.083\\                                                            & (0.036)        & (0.068)     & (0.061)      & (0.063)\\          demo\_ideology\_cLiberal                          & -0.060$^{**}$  & 0.021       & -0.024       & 0.001\\                                                            & (0.026)        & (0.037)     & (0.047)      & (0.069)\\          demo\_ideology\_cConservative+                    & -0.002         & -0.077      & -0.034       & -0.023\\                                                            & (0.024)        & (0.081)     & (0.048)      & (0.064)\\          Age                                               & -0.0008$^{**}$ & 0.001       & -0.0007      & 0.0006\\                                                            & (0.0004)       & (0.0009)    & (0.001)      & (0.0009)\\          Education                                         & 0.008          & -0.001      & 0.007        & 0.015\\                                                            & (0.006)        & (0.014)     & (0.013)      & (0.015)\\          Region: Northeast                                 & -0.009         & 0.085$^{*}$ & 0.012        & -0.035\\                                                            & (0.020)        & (0.047)     & (0.041)      & (0.041)\\          Region: Midwest                                   & -0.017         & 0.057       & 0.004        & 0.036\\                                                            & (0.019)        & (0.052)     & (0.058)      & (0.049)\\          Region: South                                     & -0.013         & 0.063       & 0.006        & -0.019\\                                                            & (0.018)        & (0.042)     & (0.035)      & (0.042)\\          Income                                            & 0.026$^{***}$  & 0.005       & 0.043$^{*}$  & 0.013\\                                                            & (0.009)        & (0.021)     & (0.023)      & (0.027)\\          Religious                                         & 0.005          & 0.044       & 0.085$^{**}$ & 0.077$^{**}$\\                                                            & (0.019)        & (0.045)     & (0.035)      & (0.035)\\          Female $\times$ demo\_ideology\_cVeryLiberal      & -0.026         & -0.134      & -0.011       & -0.069\\                                                            & (0.040)        & (0.113)     & (0.072)      & (0.073)\\          Female $\times$ demo\_ideology\_cLiberal          & 0.006          & 0.027       & -0.038       & -0.025\\                                                            & (0.032)        & (0.063)     & (0.056)      & (0.081)\\          Female $\times$ demo\_ideology\_cConservative+    & -0.056         & 0.142       & -0.108       & -0.109\\                                                            & (0.036)        & (0.114)     & (0.084)      & (0.096)\\          \midrule       \emph{Fixed-effects}\\       Questions                                         & Yes            & Yes         & Yes          & Yes\\         LLMs                                              & Yes            & Yes         & Yes          & Yes\\         \midrule       \emph{Fit statistics}\\       Observations                                      & 75,224         & 11,888      & 15,742       & 10,519\\         R$^2$                                             & 0.13495        & 0.18703     & 0.17124      & 0.15115\\         Within R$^2$                                      & 0.02051        & 0.04806     & 0.03435      & 0.03134\\         \midrule \midrule       \multicolumn{5}{l}{\emph{Clustered (case\_id) standard-errors in parentheses}}\\       \multicolumn{5}{l}{\emph{Signif. Codes: ***: 0.01, **: 0.05, *: 0.1}}\\    \end{tabular}
\end{adjustbox}
\noindent\parbox{0.8\textwidth}{\vspace{0.5em}\small\textit{Note:} Separate models estimated for each racial group, identical to Table~\ref{tab:tab_racial_subset} except that Conservative and Very Conservative are combined into a single Conservative+ category. The dependent variable is the normalized gap between LLM and human responses (LLM -- Human), where positive values indicate LLM overestimation. Reference categories: Male, Moderate (ideology), West (region). Clustered (\texttt{case\_id}) standard errors in parentheses. Fixed effects for question and LLM model included. Significance: $*** p < .01, ** p < .05, * p < .1$.}
 \end{table}

\begin{longtable}{@{}lr@{}}
\caption{\label{tab:interaction_race}Primary Model with Race Interactions}\\
\toprule
 & \multicolumn{1}{c}{LLM-Human} \\
 \midrule
\endfirsthead
\toprule
 & \multicolumn{1}{c}{LLM-Human} \\
\midrule
\endhead
\midrule
\multicolumn{2}{r}{Continued on next page}\\
\midrule
\endfoot
\bottomrule
\endlastfoot
Female & -0.045$^{***}$\\
& (0.013)\\
Age & -0.0008$^{**}$\\
& (0.0004)\\
Race/Ethnicity: Black & -0.185$^{**}$\\
& (0.092)\\
Race/Ethnicity: Other, non-Hispanic & -0.135\\
& (0.091)\\
Race/Ethnicity: Hispanic & -0.127\\
& (0.082)\\
Ideology: Very Liberal & -0.053$^{**}$\\
& (0.021)\\
Ideology: Liberal & -0.057$^{***}$\\
& (0.016)\\
Ideology: Conservative & 0.006\\
& (0.019)\\
Ideology: Very Conservative & -0.105$^{***}$\\
& (0.025)\\
Education & 0.007\\
& (0.006)\\
Income & 0.027$^{***}$\\
& (0.009)\\
Region: Northeast & -0.012\\
& (0.020)\\
Region: Midwest & -0.018\\
& (0.019)\\
Region: South & -0.014\\
& (0.018)\\
Religious & 0.010\\
& (0.019)\\
Age $\times$ Race/Ethnicity: Black & 0.001\\
& (0.0009)\\
Age $\times$ Race/Ethnicity: Other, non-Hispanic & 0.0008\\
& (0.0009)\\
Age $\times$ Race/Ethnicity: Hispanic & 0.0002\\
& (0.001)\\
Race/Ethnicity: Black $\times$ Ideology: Very Liberal & 0.106$^{*}$\\
& (0.060)\\
Race/Ethnicity: Other, non-Hispanic $\times$ Ideology: Very Liberal & 0.087$^{*}$\\
& (0.045)\\
Race/Ethnicity: Hispanic $\times$ Ideology: Very Liberal & 0.015\\
& (0.057)\\
Race/Ethnicity: Black $\times$ Ideology: Liberal & 0.095$^{***}$\\
& (0.037)\\
Race/Ethnicity: Other, non-Hispanic $\times$ Ideology: Liberal & 0.054\\
& (0.040)\\
Race/Ethnicity: Hispanic $\times$ Ideology: Liberal & 0.014\\
& (0.036)\\
Race/Ethnicity: Black $\times$ Ideology: Conservative & 0.079\\
& (0.054)\\
Race/Ethnicity: Other, non-Hispanic $\times$ Ideology: Conservative & 0.025\\
& (0.052)\\
Race/Ethnicity: Hispanic $\times$ Ideology: Conservative & -0.080\\
& (0.052)\\
Race/Ethnicity: Black $\times$ Ideology: Very Conservative & -0.208$^{**}$\\
& (0.092)\\
Race/Ethnicity: Other, non-Hispanic $\times$ Ideology: Very Conservative & -0.128\\
& (0.079)\\
Race/Ethnicity: Hispanic $\times$ Ideology: Very Conservative & 0.011\\
& (0.068)\\
Race/Ethnicity: Black $\times$ Education & -0.002\\
& (0.015)\\
Race/Ethnicity: Other, non-Hispanic $\times$ Education & -0.004\\
& (0.015)\\
Race/Ethnicity: Hispanic $\times$ Education & -0.0007\\
& (0.015)\\
Race/Ethnicity: Black $\times$ Income & -0.018\\
& (0.022)\\
Race/Ethnicity: Other, non-Hispanic $\times$ Income & -0.004\\
& (0.026)\\
Race/Ethnicity: Hispanic $\times$ Income & 0.015\\
& (0.025)\\
Race/Ethnicity: Black $\times$ Region: Northeast & 0.078$^{*}$\\
& (0.046)\\
Race/Ethnicity: Other, non-Hispanic $\times$ Region: Northeast & -0.026\\
& (0.046)\\
Race/Ethnicity: Hispanic $\times$ Region: Northeast & 0.034\\
& (0.047)\\
Race/Ethnicity: Black $\times$ Region: Midwest & 0.088\\
& (0.054)\\
Race/Ethnicity: Other, non-Hispanic $\times$ Region: Midwest & 0.076\\
& (0.049)\\
Race/Ethnicity: Hispanic $\times$ Region: Midwest & 0.016\\
& (0.064)\\
Race/Ethnicity: Black $\times$ Region: South & 0.068\\
& (0.045)\\
Race/Ethnicity: Other, non-Hispanic $\times$ Region: South & 0.014\\
& (0.043)\\
Race/Ethnicity: Hispanic $\times$ Region: South & 0.025\\
& (0.039)\\
Female $\times$ Race/Ethnicity: Black & 0.113$^{***}$\\
& (0.034)\\
Female $\times$ Race/Ethnicity: Other, non-Hispanic & 0.004\\
& (0.035)\\
Female $\times$ Race/Ethnicity: Hispanic & 0.0003\\
& (0.034)\\
Race/Ethnicity: Black $\times$ Religious & 0.080$^{*}$\\
& (0.047)\\
Race/Ethnicity: Other, non-Hispanic $\times$ Religious & 0.071$^{*}$\\
& (0.040)\\
Race/Ethnicity: Hispanic $\times$ Religious & 0.069$^{*}$\\
& (0.040)\\
\midrule
Fixed-effects\\
Questions & Yes\\
llm\_model & Yes\\
\midrule
Fit statistics\\
Observations & 113,373\\
R$^2$ & 0.15279\\
Within R$^2$ & 0.03943\\
\midrule \midrule
\multicolumn{2}{l}{Clustered (case\_id) standard-errors in parentheses}\\
\multicolumn{2}{l}{Signif. Codes: $***: p<.01, **: p<.05, *: p<.1$}\\
\end{longtable}

\section{Sensitivity Analyses}\label{app:senst}

\subsection{Single Variable Test}\label{app:senst-single}

To examine whether demographic effects on LLM prediction errors operate independently or through interactions, we conducted single-variable sensitivity tests. In these tests, LLMs received personas containing only one demographic attribute (e.g., "You are a female") rather than the full demographic profile used in the main analysis.\footnote{Religion was excluded from single-variable testing because the original survey measured religion with 15 categories (e.g., Baptist, Catholic, Jewish, Muslim, Hindu, Buddhist). Many categories had insufficient sample sizes for stable estimates, and testing each category separately would have required substantially more computational resources. We therefore operationalized religion as a binary variable (religious/not religious) in the main analysis, which was not suitable for the single-variable sensitivity test framework that examines effects across category levels.} We tested seven variables: gender, race, political ideology, income, region, education, and age. For each variable, we sampled 100 respondents from the original survey population and repeated the test five times, yielding 12,000 observations per variable (100 respondents $\times$ 20 questions $\times$ 6 models $\times$ 5 replications / 5 replications = 12,000).

We compared these isolated effects to the controlled effects from our main full-persona regression ($N$ = 113,373). The isolated effect represents the LLM-human gap difference when only one variable is known, while the controlled effect represents the same difference when all demographic variables are included simultaneously.

\begin{table}[H]
\caption{\label{tab:sensitivity}Isolated vs. Controlled Effects: Single-Variable Sensitivity Analysis}
\begin{adjustbox}{center}
\begin{tabular}{llrrrc}
\toprule
\multicolumn{2}{c}{ } & \multicolumn{3}{c}{Effect Size} & \multicolumn{1}{c}{ } \\
\cmidrule(l{3pt}r{3pt}){3-5}
Variable & Comparison & Isolated & Controlled & Difference & Sign Reversal\\
\midrule
Gender & Female - Male & 0.135 & -0.030 & 0.165 & Yes\\
Race & Black - White & -0.076 & 0.046 & -0.122 & Yes\\
Political & Liberal - Conservative & -0.285 & -0.047 & -0.238 & No\\
Income & Low - High & -0.115 & -0.025 & -0.090 & No\\
Region & South - Northeast & 0.059 & 0.002 & 0.057 & No\\
\addlinespace
Education & Low - High & 0.195 & -0.007 & 0.202 & Yes\\
Age & Correlation & -0.060 & -0.001 & -0.059 & No\\
\bottomrule
\end{tabular}   
\end{adjustbox}
\noindent\parbox{0.8\textwidth}{\vspace{0.5em}\small\textit{Note:} Isolated effects from single-variable persona tests ($N$ = 12{,}000 each). Controlled effects from full-persona regression with demographic controls and question/model fixed effects.}
\end{table}

Table~\ref{tab:sensitivity} and Figure~\ref{fig:sensitivity} present the results. Four of seven variables show sign reversals between isolated and controlled specifications: gender, race, region, and education. Most notably, gender and race---the variables central to our racialized gender stereotype finding---both reverse direction.

\begin{figure}[H]
    \centering
    \includegraphics[width=0.8\textwidth]{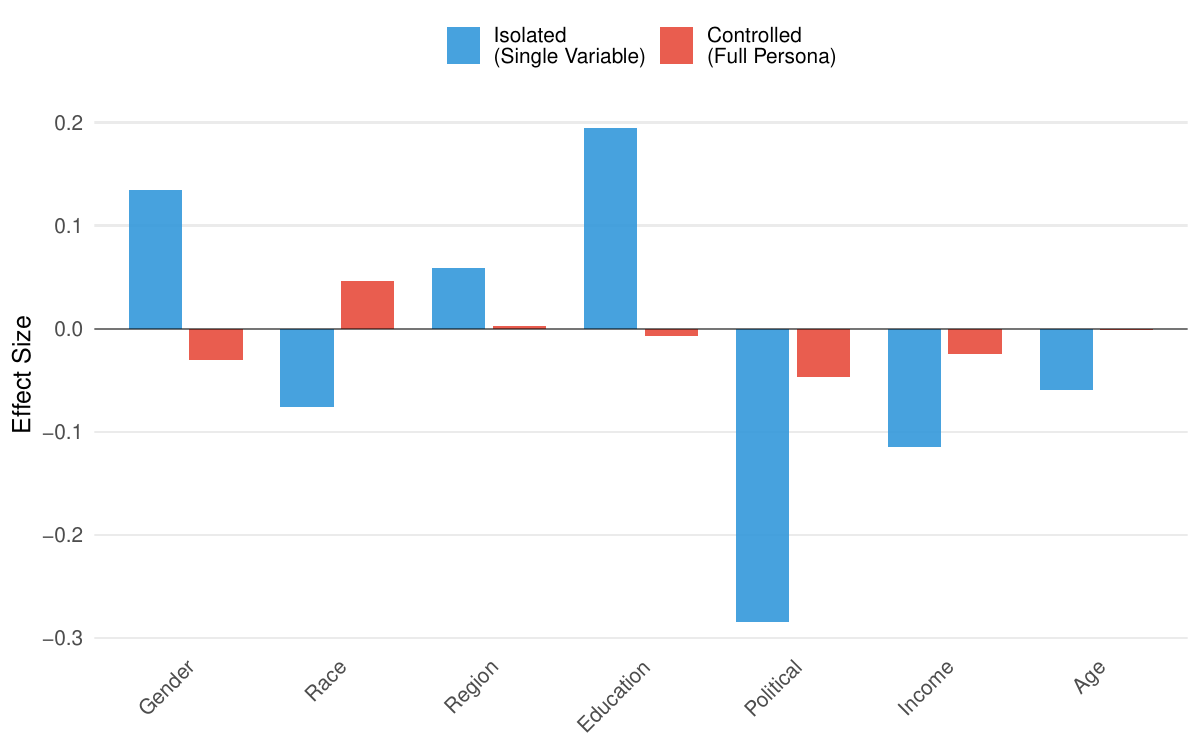}
    \caption{\titlecap{Comparison of isolated and controlled demographic effects on LLM prediction errors}. Isolated effects (blue) are calculated from single-variable persona tests where LLMs received only one demographic attribute ($N \approx 12{,}000$ per variable). Controlled effects (red) are extracted from full-persona regression with all demographic controls ($N$ = 113,373). Variables are ordered by whether they show sign reversal: Gender, Race, Region, and Education reverse direction between specifications, while Political, Income, and Age maintain the same direction but with attenuated magnitude. Comparisons: Gender (Female $-$ Male), Race (Black $-$ White), Region (South $-$ Northeast), Political (Liberal $-$ Conservative), Income (Low $-$ High), Education (Low $-$ High), Age (correlation coefficient). Positive values indicate greater LLM overestimation for the first group in each comparison.}
    \label{fig:sensitivity}
\end{figure}

For gender, the isolated effect is positive (+0.135), suggesting LLMs overestimate female climate opinions when gender is the only known attribute. However, the controlled effect is negative ($-$0.030), indicating LLMs underestimate female opinions when the full persona is known. This reversal occurs because the gender effect depends on race: LLMs underestimate White women relative to White men but overestimate Black women relative to Black men. When only gender is provided, LLMs cannot capture this interaction.

Similarly, for race, the isolated effect is negative ($-$0.076), suggesting LLMs underestimate Black respondents when race is the only known attribute. The controlled effect is positive (+0.046), indicating overestimation when the full persona is known. This reversal reflects the gender dependence: the models overestimate Black respondents overall, and Black women more than Black men (Table~\ref{tab:gender_by_race}). The isolated test, lacking gender information, misses this pattern.

Region and education also show sign reversals, though these are not driven by clear two-way interactions with race or gender. The mechanisms underlying these reversals may involve higher-order interactions or compositional assumptions that LLMs make when given minimal demographic information.

Three variables---political ideology, income, and age---do not show sign reversals. However, the isolated effects are substantially larger in magnitude than controlled effects (7-100 times larger), suggesting that full-persona context attenuates these effects even when it does not reverse them.

These findings demonstrate that single-variable sensitivity tests can yield misleading conclusions about LLM demographic biases. The key insight from our main analysis—that LLMs exhibit racialized gender stereotypes—is invisible in isolated tests because it emerges through the interaction between gender and race. This underscores the importance of testing LLMs with realistic, multidimensional demographic profiles rather than isolated attributes.

\subsection{Saturation Analysis}\label{app:senst-saturation}

To further examine how persona complexity affects LLM prediction errors, we conducted a saturation analysis that progressively removes variables from the full persona. This approach reveals whether demographic effects are stable across different levels of persona detail or emerge only with sufficient contextual information.

We constructed six saturation levels by sequentially removing variables from the full seven-variable persona: (1) all variables except political ideology; (2) all except political ideology and region; (3) all except political, region, and education; (4) gender, income, and age only; (5) gender and income only; and (6) single variable only (from the sensitivity analysis). The removal order prioritized political ideology first, as prior research identifies it as the strongest predictor of climate opinions \citep{hornseyMetaanalysesDeterminantsOutcomes2016, dunlapWideningGapRepublican2008}.\footnote{We focus on gender and race effects because they are central to our main finding of racialized gender stereotypes. Other variables (political ideology, income, region, education, age) either have insufficient measurement points due to the removal sequence (e.g., political ideology appears only at levels 1 and 7) or are not central to our theoretical contribution. The saturation design prioritizes testing whether the gender-race interaction pattern holds across different levels of persona complexity, rather than tracking all demographic effects.}

\vspace{1em}
\begin{table}[H]
\caption{\label{tab:saturation}Saturation Analysis: Gender and Race Effects by Number of Persona Variables}
\begin{adjustbox}{center}
\begin{tabular}[t]{clrrr}
\toprule
\multicolumn{2}{c}{ } & \multicolumn{2}{c}{Effect Size} & \multicolumn{1}{c}{ } \\
\cmidrule(l{3pt}r{3pt}){3-4}
N Variables & Variables Included & Gender & Race & N\\
\midrule
7 & All & -0.115 & 0.156 & 113,373\\
6 & All except political & -0.135 & -0.02 & 59,342\\
5 & All except political, region & -0.129 & -0.01 & 59,176\\
4 & All except political, region, education & -0.118 & 0.003 & 59,147\\
3 & Gender, income, age & -0.130 & --- & 59,776\\
\addlinespace
2 & Gender, income & -0.143 & --- & 59,692\\
1 & Single variable only & 0.135 & -0.076 & 11,900 /  11,537\\
\bottomrule
\end{tabular}
\end{adjustbox}
\noindent\parbox{0.8\textwidth}{\vspace{0.5em}\small\textit{Note:} Gender effect: Female $-$ Male. Race effect: Black $-$ White. Race effect not estimable at levels 2-3 because race was not included in the persona. Positive values indicate greater LLM overestimation for the first group. Effect sizes are computed on the raw (unscaled) response gap ($\texttt{llm\_response\_numeric} - \texttt{human\_response}$) rather than the min--max--scaled gap used in Table~\ref{tab:main}, so magnitudes are not directly comparable across the two. Level 7 uses full-persona data ($N$ = 113,373). Levels 2-6 use saturation test data ($N$ $\approx$ 60,000 each). Level 1 uses single-variable test data ($N$ shown as Gender / Race).}
\end{table}

Table~\ref{tab:saturation} and Figure~\ref{fig:saturation} present the results. The gender effect shows remarkable stability across levels 2--7, ranging from $-0.117$ to $-0.143$ (all negative, indicating underestimation of female climate opinions). The reversal to $+0.135$ occurs only at level 1, when gender is completely isolated. This pattern suggests that LLMs apply a fundamentally different stereotype when given minimal information versus any degree of contextual detail.

\begin{figure}[H]
\centering
\includegraphics[width=0.8\textwidth]{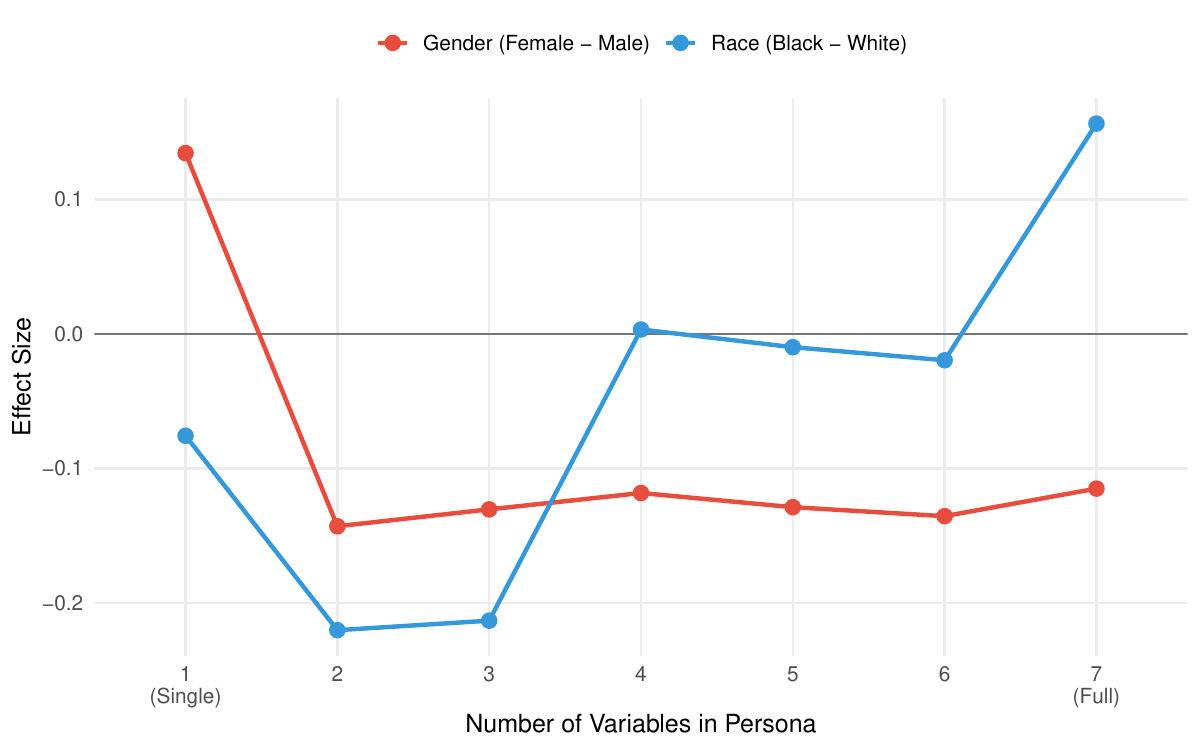}
\caption{\titlecap{Gender and race effects across saturation levels}. The x-axis shows the number of variables included in the LLM persona, from 1 (single variable) to 7 (full persona). Gender effect (red): Female $-$ Male gap difference. Race effect (blue): Black $-$ White gap difference. Race effect is not estimable at levels 2--3 because race was not included in those persona configurations. The gender effect remains stable and negative across levels 2--7, reversing only at level 1. The race effect is positive at level 7, near zero at levels 4--6, and negative at level 1. The dashed horizontal line indicates zero (no effect).}
\label{fig:saturation}
\end{figure}

The race effect shows a different pattern. At the full-persona level (7 variables), LLMs overestimate Black respondents relative to White respondents ($+0.148$). However, this effect diminishes substantially at levels 4--6 (near zero: $-0.020$ to $+0.003$), and reverses at level 1 ($-0.076$, underestimation). This instability suggests that the race effect is more sensitive to persona composition than the gender effect.

These findings have two implications. First, single-variable tests produce misleading conclusions---the gender effect reverses sign, and the race effect changes substantially. Second, the gender-race interaction documented in our main analysis is not an artifact of including too many variables; rather, the consistent negative gender effect across levels 2--7 suggests that even minimal context (two variables) is sufficient to produce the direction of bias observed in full personas. The sharp reversal at level 1 indicates that LLMs behave qualitatively differently when given isolated demographic attributes versus any combination of attributes.

\subsection{Temperature Sensitivity Analysis}\label{app:temperature}
To validate our choice of temperature = 0 for the main analysis, we tested whether temperature settings affect LLM responses to climate opinion questions. We queried each model with baseline prompts (no demographic persona) at three temperature levels: 0.1, 0.5, and 1.0. Baseline conditions were used because temperature effects should operate at the model level, independent of persona content; temperature-insensitivity at baseline implies similar stability under persona conditions.

\begin{table}[H]
\caption{\label{tab:temp_sensitivity}Baseline Statistics by Model and Temperature}
\begin{adjustbox}{center}
\begin{tabular}[t]{lrrrr}
\toprule
llm\_model & temperature & baseline\_mean & baseline\_sd & n\\
\midrule
gemma3 & 0.1 & 0.692 & 0.168 & 2000\\
gemma3 & 0.5 & 0.692 & 0.168 & 2000\\
gemma3 & 1.0 & 0.692 & 0.168 & 2000\\
gpt-4o-2024-08-06 & 0.1 & 0.956 & 0.109 & 2000\\
gpt-4o-2024-08-06 & 0.5 & 0.956 & 0.109 & 2000\\
\addlinespace
gpt-4o-2024-08-06 & 1.0 & 0.957 & 0.113 & 2000\\
gpt-5-2025-08-07 & 1.0 & 0.973 & 0.096 & 600\\
llama3 & 0.1 & 0.745 & 0.141 & 2000\\
llama3 & 0.5 & 0.745 & 0.141 & 2000\\
llama3 & 1.0 & 0.745 & 0.141 & 2000\\
\addlinespace
mistral & 0.1 & 0.947 & 0.153 & 2000\\
mistral & 0.5 & 0.947 & 0.153 & 2000\\
mistral & 1.0 & 0.947 & 0.153 & 2000\\
phi4 & 0.1 & 0.922 & 0.123 & 2000\\
phi4 & 0.5 & 0.922 & 0.123 & 2000\\
\addlinespace
phi4 & 1.0 & 0.922 & 0.123 & 2000\\
\bottomrule
\end{tabular}
\end{adjustbox}
\end{table}

Results show virtually no effect of temperature on either mean responses or response variance (Table~\ref{tab:temp_sensitivity}). All models produced nearly identical outputs across temperature settings, confirming that our conservative temperature setting (0) does not constrain the findings. GPT-5 was tested only at temperature = 1.0, as this chain-of-thought model does not permit lower temperature settings; its response patterns were comparable to other models at their default configurations.

\subsection{Top\_p Sensitivity Analysis}\label{app:top_p}

\begin{table}[H]
\caption{\label{tab:baseline_by_top_p}Baseline Statistics by Model and Top\_p}
\begin{adjustbox}{center}
\begin{tabular}[t]{lrrrr}
\toprule
llm\_model & top\_p & baseline\_mean & baseline\_sd & n\\
\midrule
gemma3 & 0.1 & 0.692 & 0.168 & 2000\\
gemma3 & 0.5 & 0.692 & 0.168 & 2000\\
gemma3 & 0.9 & 0.692 & 0.168 & 2000\\
gpt-4o-2024-08-06 & 0.1 & 0.948 & 0.116 & 2000\\
gpt-4o-2024-08-06 & 0.5 & 0.949 & 0.115 & 2000\\
\addlinespace
gpt-4o-2024-08-06 & 0.9 & 0.950 & 0.115 & 2000\\
llama3 & 0.1 & 0.745 & 0.141 & 2000\\
llama3 & 0.5 & 0.745 & 0.141 & 2000\\
llama3 & 0.9 & 0.745 & 0.141 & 2000\\
mistral & 0.1 & 0.947 & 0.153 & 2000\\
\addlinespace
mistral & 0.5 & 0.947 & 0.153 & 1999\\
mistral & 0.9 & 0.947 & 0.153 & 2000\\
phi4 & 0.1 & 0.922 & 0.123 & 2000\\
phi4 & 0.5 & 0.922 & 0.123 & 2000\\
phi4 & 0.9 & 0.922 & 0.123 & 2000\\
\bottomrule
\end{tabular}
\end{adjustbox}
\end{table}

\subsection{Decoding Sensitivity Under Persona Conditioning}\label{app:decoding}

{
The sensitivity analyses reported in Appendices~\ref{app:temperature} and \ref{app:top_p} used baseline prompts without demographic personas, leaving open whether decoding parameters behave the same way once a persona is attached. We therefore repeated the test under persona conditioning. From the 978 analytic respondents we drew a stratified subset of 76 personas covering all 40 race $\times$ gender $\times$ ideology cells, including every cell the intersectional analysis reports. Each model answered all 20 questions for all 76 personas under three decoding configurations, three times each: setting A as used for the main results, setting B holding temperature at 0 while moving top\_p to the opposite end of its range, and setting C at temperature 1 with top\_p 1. The design yields 48 invocations of 1,520 responses each. GPT-5 ran setting C only, since the API coerces its temperature to 1 regardless of what is requested.

Top\_p has no effect at temperature 0. For all four locally served models, settings A and B produced byte-identical output across all three repeats, at every persona and every item, despite top\_p differing by a factor of nine between them. The claim that both sampling parameters are functionally inert for single-token numeric responses is therefore not an inference from the concentration of the probability distribution but a direct observation: the parameter can be moved across its usable range without changing a single response.

Determinism at temperature 0 is not universal. All four locally served models returned identical responses across all three repeats under both temperature-0 settings. GPT-4o did not, agreeing on 91.0\% and 92.4\% of persona-item cells under settings A and B. Setting temperature to 0 therefore does not guarantee reproducibility through a hosted API, where batching, load balancing, and server-side implementation details can vary between calls in ways a client cannot observe or control. The difference is one of access mode rather than of model size or architecture, and it is invisible without repeated querying.

Temperature is the parameter that matters, and models diverge sharply under it. At temperature 1, agreement across repeats ranged from 43.6\% for Llama-3 to 89.3\% for Gemma-3 (Table~\ref{tab:decoding}). The spread indicates that these models differ considerably in how concentrated their output distributions are for structured survey items, so a study sampling at temperature 1 would obtain different levels of within-model consistency depending on which models it included.

Sampling at temperature 1 also changes what a compression analysis would measure. Measured compression is a variance ratio, so noise from the decoder enters the numerator: pooled across the five models testable at both settings, mean measured compression falls from 44.4\% at temperature 0 to 40.8\% at temperature 1, and the size of the change is proportional to each model's self-consistency (Table~\ref{tab:decoding}). Llama-3, the least self-consistent, moves 14.2 percentage points, while the four models that remain largely self-consistent move 2.0 points or less. The levels are not comparable to the compression figures in Section~\ref{subsec:compression}, since the 76 personas were stratified across all 40 cells rather than drawn to be representative, so only the contrast between settings is interpretable.

GPT-4o's instability could in principle concentrate in particular subgroups, so we checked whether it falls disproportionately on the smallest cells, where a single altered response would move an estimate. It does not. Across the five Black very conservative personas in the subset, the 100 persona-item cells agreed at 92.0\% for GPT-4o under setting A, against 91.0\% across all 76 personas, and all four locally served models were perfectly consistent there. Under stochastic decoding the same cell degrades faster than the subset as a whole, falling to 30.0\% for Llama-3 against 43.6\% overall and 37.4\% for Phi-4 against 55.7\%, so the deterministic configuration matters most exactly where the estimates are thinnest. What remains at issue for the race-specific estimates in Table~\ref{tab:tab_racial_subset} is therefore the number of respondents behind them, which Appendix~\ref{app:smallcell} takes up. \begin{table}[H]
\caption{Response Agreement and Measured Compression Under Persona Conditioning}\label{tab:decoding}
\begin{adjustbox}{center}
\begin{tabular}[t]{lrrrrr}
\toprule
 & \multicolumn{3}{c}{Response agreement} & \multicolumn{2}{c}{Measured compression} \\
\cmidrule(lr){2-4} \cmidrule(lr){5-6}
 & A (temp 0) & B (temp 0) & C (temp 1) & A (temp 0) & C (temp 1) \\
\midrule
Llama-3 & 100.0\% & 100.0\% & 43.6\% & 84.6\% & 70.5\% \\
Gemma-3 & 100.0\% & 100.0\% & 89.3\% & 43.8\% & 43.7\% \\
Phi-4 & 100.0\% & 100.0\% & 55.7\% & 30.5\% & 28.5\% \\
Mistral & 100.0\% & 100.0\% & 79.3\% & 29.3\% & 28.2\% \\
GPT-4o & 91.0\% & 92.4\% & 70.4\% & 33.6\% & 33.1\% \\
GPT-5 & --- & --- & 85.1\% & --- & 34.6\% \\
\bottomrule
\end{tabular}
\end{adjustbox}
\noindent\parbox{0.8\textwidth}{\vspace{0.5em}\small\textit{Note:} Response agreement is the percentage of (persona, item) cells returning an identical response across three repeats of the same configuration, computed over cells where all three returned a response. Measured compression is one minus the mean per-question ratio of model variance to human variance, averaged over the three repeats. Setting A is the configuration used for the main results; setting B holds temperature at 0 and moves top\_p to the opposite end of its range; setting C samples at temperature 1. GPT-5 ran setting C only, since the API coerces its temperature, and returned parseable responses for 93.4--94.3\% of queries across its three repeats. Compression levels are not comparable to the compression figures in Section~\ref{subsec:compression}: these 76 personas were stratified across all 40 race by gender by ideology cells rather than drawn to be representative, so only the contrast between settings is interpretable.}
\end{table}
}

\subsection{Sensitivity to the Timing Item}\label{app:missingness}

{
The timing item---``When do you think global warming will start to harm people in the United States?'' (\texttt{when\_harm\_US})---is distinctive in two respects. It offers six response options, more than any other item, and its lowest, ``Never,'' is the only response in the instrument that denies climate harm outright. Selection of that option rises monotonically with conservatism among human respondents and among the models predicting them (Table~\ref{tab:never}). The models choose it less often than the respondents they simulate at every level above very liberal: 10.2\% against 27.2\% among conservatives, and 29.4\% against 53.9\% among very conservatives. The same compression toward the population mean reported in Section~\ref{subsec:compression} is visible here within a single item.

Models differ in how readily they select it. Gemma-3 chose ``Never'' on 29.9\% of its responses to this item, against 5.8\% for Mistral, 1.1\% for Phi-4, 0.3\% for Llama-3, 0.1\% for GPT-4o, and none for GPT-5. Within Gemma-3 the gradient runs from 2.5\% among very liberal personas to 95.5\% among very conservative personas, steeper than the human gradient of 0.0\% to 53.9\%, so on this item that model conditions on ideology more strongly than the respondents it simulates. Three of the six selected the option in fewer than 1\% of responses. Because the item is distinctive in both range and ideological structure, we re-estimated the main analyses without it (Table~\ref{tab:robust_timing}). The conclusions do not depend on it. The mean per-question variance ratio moves from 0.706 to 0.709, so predicted opinions remain substantially compressed. The pooled gender effect holds at $\beta = -0.029$ ($p < .01$) against $-0.030$ in the full specification. The two race-specific interactions persist, positive for Black respondents ($\beta = +0.335$, $p < .01$, against $+0.323$) and negative for Hispanic respondents ($\beta = -0.305$, $p < .01$, against $-0.319$). \begin{table}[H]
\caption{Selection of ``Never'' on the Timing Item by Persona Ideology}\label{tab:never}
\begin{adjustbox}{center}
\begin{tabular}[t]{lrr}
\toprule
Persona ideology & Human respondents & LLM predictions \\
\midrule
Very liberal & 0.0\% & 0.4\% \\
Liberal & 3.0\% & 0.8\% \\
Moderate & 9.9\% & 2.9\% \\
Conservative & 27.2\% & 10.3\% \\
Very conservative & 53.9\% & 29.4\% \\
\bottomrule
\end{tabular}
\end{adjustbox}
\noindent\parbox{0.8\textwidth}{\vspace{0.5em}\small\textit{Note:} Percentages are the share selecting ``Never'' in response to when global warming will begin harming people in the United States, computed on valid responses to that item in the analytic sample. Human $n = 975$; LLM $n = 5{,}849$, pooling six models. Across the six the share runs from 0.0\% to 29.9\%.}
\end{table}
}

\begin{table}[H]

\caption{Robustness of Gender and Intersectional Effects to Excluding the Timing Item}\label{tab:robust_timing}
\begin{adjustbox}{width=\textwidth, center}
\begin{tabular}{lcccccc}       \tabularnewline \midrule \midrule                                                    & Pooled: all 20 & Pooled: excl.  & Black: all 20 & Black: excl.  & Hispanic: all 20 & Hispanic: excl. \\          Model:                                       & (1)            & (2)            & (3)           & (4)           & (5)              & (6)\\         \midrule       \emph{Variables}\\       Female                                       & -0.030$^{***}$ & -0.029$^{***}$ & 0.069         & 0.080$^{*}$   & -0.008           & -0.008\\                                                       & (0.011)        & (0.011)        & (0.043)       & (0.043)       & (0.041)          & (0.040)\\          Female $\times$ Ideology: Very Liberal       &                &                & -0.133        & -0.149        & -0.012           & -0.031\\                                                       &                &                & (0.111)       & (0.112)       & (0.070)          & (0.070)\\          Female $\times$ Ideology: Liberal            &                &                & 0.015         & 0.0006        & -0.041           & -0.038\\                                                       &                &                & (0.062)       & (0.061)       & (0.056)          & (0.054)\\          Female $\times$ Ideology: Conservative       &                &                & -0.014        & -0.041        & -0.009           & -0.004\\                                                       &                &                & (0.103)       & (0.100)       & (0.100)          & (0.098)\\          Female $\times$ Ideology: Very Conservative  &                &                & 0.323$^{***}$ & 0.335$^{***}$ & -0.319$^{***}$   & -0.305$^{***}$\\                                                       &                &                & (0.111)       & (0.110)       & (0.116)          & (0.117)\\          \midrule       \emph{Fixed-effects}\\       Questions                                    & Yes            & Yes            & Yes           & Yes           & Yes              & Yes\\         LLMs                                         & Yes            & Yes            & Yes           & Yes           & Yes              & Yes\\         \midrule       \emph{Fit statistics}\\       Observations                                 & 113,373        & 107,524        & 11,888        & 11,265        & 15,742           & 14,932\\         R$^2$                                        & 0.14068        & 0.14383        & 0.22931       & 0.23910       & 0.18257          & 0.19026\\         Within R$^2$                                 & 0.02570        & 0.02656        & 0.09756       & 0.10305       & 0.04756          & 0.04861\\         \midrule \midrule       \multicolumn{7}{l}{\emph{Clustered (case\_id) standard-errors in parentheses}}\\       \multicolumn{7}{l}{\emph{Signif. Codes: ***: 0.01, **: 0.05, *: 0.1}}\\    \end{tabular}
\end{adjustbox}
\noindent\parbox{\textwidth}{\vspace{0.5em}\small\textit{Note:} Columns (1)--(2) are the pooled model; (3)--(6) are race-stratified models with Gender $\times$ Ideology interactions. Odd columns use all 20 items; even columns exclude the timing item (\texttt{when\_harm\_US}) for all models. The dependent variable is the normalized LLM--human gap, where positive values indicate LLM overestimation. Reference categories: Male, Moderate (ideology). Clustered (\texttt{case\_id}) standard errors in parentheses. Fixed effects for question and LLM model included. Significance: $*** p < .01, ** p < .05, * p < .1$.}

\end{table}

\subsection{Robustness of the Race-Specific Interactions at Small Cell Sizes}\label{app:smallcell}

{
The race-specific Female~$\times$~Very~Conservative interactions in Table~\ref{tab:tab_racial_subset} rest on small cells, and we report here what survives when those estimates are tested as harshly as the data allow. Five Black and twelve Hispanic respondents identify as very conservative, and the Black cell contains a single woman (Table~\ref{tab:cell_sizes}). The responses underlying these estimates are reproducible under the configuration used for the main results (Appendix~\ref{app:decoding}), so what follows concerns the number of respondents rather than the stability of their predicted answers. Four checks follow: a specification that widens the ideological band, a correction for multiple comparisons, exact permutation tests at the smallest cells, and partial pooling across the full race by ideology grid. Where a check admits more than one defensible specification, we report the more conservative one. A difference in sign between Black and Hispanic respondents survives every check; no claim about the magnitude of either does.

Collapsing conservatives with very conservatives tests whether the smallest cell drives the result. The recoded band raises the Black subgroup from 5 respondents to 18 (11 men, 7 women) and the Hispanic subgroup from 12 to 38, and we refit the four race-stratified models with no other change: the same controls, the same question and model fixed effects, the same clustering on \texttt{case\_id}, and Moderate as the reference category (Table~\ref{tab:tab_racial_subset_collapsed}). Both signs hold. The Black interaction remains positive ($\beta=+0.142$, $SE=0.114$) and the Hispanic interaction negative ($\beta=-0.108$, $SE=0.084$), while neither reaches conventional significance in the wider band ($p=.214$ and $p=.204$). Widening the band therefore separates two questions that the extreme-band specification answered together: the direction of the race-specific difference does not depend on the single-woman cell, and its statistical significance does.

We correct for multiple comparisons across the four race-stratified interactions. Testing the same interaction in four racial groups raises the chance that one of them clears conventional significance by chance alone, so the uncorrected $p$-values overstate what four tests can establish. Treating the four race-stratified Female~$\times$~Very~Conservative interactions as a single family and targeting a false discovery rate of .05, the Benjamini--Hochberg procedure \citep{benjaminiControllingFalseDiscovery1995} retains both the Black and Hispanic estimates, whose adjusted $p$-values ($q$) are .014 in each case (Table~\ref{tab:bh}). The White interaction, reported at $p<.1$ in the main text, does not survive ($q=.100$). The choice of family is itself a judgment, so we applied the same procedure to the wider set of all sixteen Female~$\times$~Ideology interactions estimated across the four models, where both estimates fall just outside the same threshold ($q=.054$). The result does not turn on the choice of correction. Bonferroni, which bounds the familywise error rate rather than the false discovery rate, also retains both estimates in the four-test family (adjusted $p=.018$ for Black and $p=.027$ for Hispanic) and retains neither in the sixteen-test family ($p=.072$ and $p=.108$). The four race-stratified interactions are the family the argument defines, since each one tests the gender contrast at the ideological extreme within a single racial group. Correcting for multiplicity removes one of the four race-specific claims and leaves the other two standing, at a margin that is comfortable in the four-test family and thin in a family four times larger. \begin{table}[H]
\caption{\label{tab:bh}Benjamini--Hochberg Correction Across the Four Race-Stratified Female~$\times$~Very~Conservative Interactions}
\begin{adjustbox}{center}
\begin{tabular}[t]{lrrrr}
\toprule
Race & $\beta$ & $p$ & BH threshold & $q$ \\
\midrule
Black    & $+0.323$ & .005 & .013 & \textbf{.014} \\
Hispanic & $-0.319$ & .007 & .025 & \textbf{.014} \\
White    & $-0.096$ & .075 & .038 & .100 \\
Other    & $-0.097$ & .496 & .050 & .496 \\
\bottomrule
\end{tabular}
\end{adjustbox}
\noindent\parbox{0.8\textwidth}{\vspace{0.5em}\small\textit{Note:} Tests ordered by $p$. The threshold for the $k$th of $m$ ordered tests is $(k/m)\alpha$ with $\alpha=.05$; $q$ reports false-discovery-rate adjusted $p$-values. Bold entries survive at $q<.05$. Applying the same procedure to the wider family of all 16 Female~$\times$~Ideology interactions estimated across the four models returns $q=.054$ for both the Black and Hispanic estimates, just outside the same threshold.}
\end{table}

Exact permutation tests bound what inference the smallest cells can support at all. Reassigning the gender labels within the Black very conservative cell yields only five distinct arrangements, since the cell holds four men and one woman. The observed difference is the largest of the five ($+0.413$; the others are $+0.230$, $-0.064$, $-0.254$, and $-0.324$), giving a one-sided exact $p$ of $1/5=.200$. That figure is a floor rather than a result: no test applied to five respondents can return a smaller one. The regression coefficient for this group is estimated across the full Black subsample with fixed effects and clustered errors, so it is not the same quantity as the cell mean difference, but it draws its identification at the very conservative level from the same five respondents. The Hispanic cell behaves differently. Seven men and five women admit 792 arrangements, and the observed difference of $-0.339$ falls at exact $p=.0025$, well inside a floor of $.0013$. We therefore report the two cells as asymmetric in what they can bear: the Hispanic estimate rests on assumption-free evidence, the Black estimate does not.

Partial pooling shrinks each cell toward the grand mean in proportion to its imprecision, and we apply it across the full grid rather than only the very conservative row. Treating the twenty race by ideology gender gaps as draws from a common distribution, with a pooled within-cell variance and a method-of-moments estimate of the between-cell variance, the Black very conservative estimate receives a weight of $0.22$ and shrinks from $+0.413$ to $+0.075$, a reduction of 82 percent. The Hispanic estimate, resting on twelve respondents, shrinks less, from $-0.339$ to $-0.184$. Pooling over the four very conservative cells alone would shrink the Black estimate only to $+0.283$; we report the grid-wide figure because it borrows strength from more cells and therefore penalizes the smallest one harder. After shrinkage the Black very conservative cell remains the only positive gap among the four races, and Black respondents remain the only racial group with a positive gap at any ideology level, holding at four of the five while White, Hispanic, and Other respondents are negative in all fifteen of theirs, which places the reversal within a pattern running the length of the row rather than in isolation.

We did not bootstrap the smallest cell. Resampling respondents with replacement inside a cell holding one woman returns that same woman in every draw, so the resulting interval reflects variation among the four men and nothing else. Run naively, the procedure returns $+0.413$ with a 95 percent interval of $[+0.243, +0.562]$: narrow, excluding zero, and more confident than five observations can warrant. Bootstrapping is sound in general and degenerates in this case, so we use exact permutation and partial pooling instead, both of which remain well defined at $n=1$.

Taken together, the four checks support a narrow claim and refuse a wider one: the sign of the race-specific difference holds under a wider ideological band, survives correction for multiple comparisons across the four race-stratified tests while falling just outside it under a family of sixteen, and persists after shrinkage, while its magnitude rests on cells too small to estimate reliably and should not be read as a quantity.
} \subsection{Abstention Under a Matched Response Space}\label{app:matched}

{
Comparing abstention between humans and models requires both to have the option available, and two items meet that condition. As Section~\ref{subsec:measures} sets out, nine items presented a don't-know category to the models, but on six of them the codebook assigns it the value 0, which the system prompt excluded. Of the three remaining, the cause item drew no human abstentions at all, leaving whether global warming is happening (category 2, ``Don't know'') and scientific consensus (category 1, ``Don't know enough to say''). On both, humans and models chose from the same numbered set, so a difference in abstention rate reflects behavior rather than differing option sets.

On both items the models abstained less often than the respondents they were simulating. Human respondents selected ``Don't know'' on the happening item at 12.6\% (123 of 977) against 8.5\% for the models (487 of 5,723), and on scientific consensus at 16.2\% (158 of 977) against 0.7\% (43 of 5,837). The gap on the consensus item is more than twentyfold. Where an abstention option was available to both, the models did not over-abstain; they under-abstained, expressing a settled view on questions where roughly one human respondent in six declined to. The pattern is consistent with the variance compression reported in Section~\ref{subsec:compression}: a model that rarely admits uncertainty is the same model that rarely reaches the ends of a scale.

Abstention behavior is not a property of language models in general but of particular ones (Table~\ref{tab:abstention}). Only Llama-3 never abstained on either item, and three of the six never abstained on the happening item. Gemma-3 abstained on more than a third of the happening item and never on the consensus item, while GPT-5 abstained at moderate rates on both. Gemma-3 is also the model whose responses vary most with the persona on the timing item (Appendix~\ref{app:missingness}), so the model that abstains most readily here conditions most strongly there.

Two caveats bound what the comparison establishes. It rests on two of twenty items, both measuring belief about fact rather than attitude or policy preference, so it does not speak to abstention on the other eighteen. The models also answered under an instruction to return a single number, which may suppress abstention independently of whether the category was listed. \begin{table}[H]
\caption{Abstention Rates by Respondent Type and Model on the Two Items With Matched Response Options}\label{tab:abstention}
\begin{adjustbox}{center}
\begin{tabular}[t]{lrr}
\toprule
 & Happening (cat.\ 2) & Sci.\ consensus (cat.\ 1) \\
\midrule
Human respondents & 12.6\% (123/977) & 16.2\% (158/977) \\
\midrule
Gemma-3 & 34.6\% (334/964) & 0.0\% (0/964) \\
GPT-5 & 13.5\% (115/854) & 3.6\% (35/966) \\
GPT-4o & 3.9\% (38/977) & 0.1\% (1/977) \\
Llama-3 & 0.0\% (0/977) & 0.0\% (0/977) \\
Mistral & 0.0\% (0/974) & 0.6\% (6/976) \\
Phi-4 & 0.0\% (0/977) & 0.1\% (1/977) \\
\midrule
All models & 8.5\% (487/5,723) & 0.7\% (43/5,837) \\
\bottomrule
\end{tabular}
\end{adjustbox}
\noindent\parbox{0.8\textwidth}{\vspace{0.5em}\small\textit{Note:} Each cell gives the share of valid responses in the analytic sample that selected the don't-know category, with the count and denominator in parentheses. Denominators differ across models because models returned different numbers of parseable responses. The eighteen remaining items either present no such category or code it outside the range the system prompt permitted, so they cannot support this comparison.}
\end{table}
}

\subsection{Robustness to Don't-Know Handling}\label{app:dkrobust}

{
The CCAM instrument codes its ``don't know'' category inconsistently across items. On the five harm items and the media-exposure item it sits at 0, outside the substantive response scale. On two items it sits inside the response ordering: code 2 of three on whether global warming is happening, placing the respondent between ``No'' and ``Yes,'' and code 1 of four on scientific consensus. Our primary specification follows the instrument, dropping the category where it falls outside the scale and retaining it at its coded position where the instrument places it within the ordering, which keeps 281 human responses carrying ordinal information.

To test whether that treatment matters, we re-estimated the analysis with don't-know responses dropped on every item offering one. The analytic sample falls from 113,373 to 111,221 responses, the mean prediction gap moves from 0.060 to 0.059, and mean variance compression from 29.4\% to 28.5\%. The compression result does not depend on how the category is handled.

The race-specific interactions retain their direction under the same specification. The Female~$\times$~Very~Conservative interaction remains positive for Black respondents ($\beta = +0.258$, $SE = 0.127$, $p = .044$) and negative for Hispanic respondents ($\beta = -0.370$, $SE = 0.119$, $p = .002$), with White and Other respondents negative and not significant, as in the primary specification. Each estimate remains individually significant where it was. Benjamini--Hochberg correction across the four race-stratified tests separates them in this specification, however: the Hispanic estimate survives at $q = .009$ while the Black estimate does not, at $q = .089$ against a target of .05, although both survived in the primary specification. The direction of the race-specific difference, which is what we claim, holds under either treatment of the category.
}


\twocolumn
\bibliographystyle{cas-model2-names}

\bibliography{library_dedup}



\end{document}